\documentclass{aa}  
\usepackage{graphicx}

\usepackage{subcaption}  
\usepackage{comment}

\usepackage{txfonts}

\usepackage[usenames,dvipsnames]{color} 

\usepackage{booktabs}  
\usepackage{mathtools} 

\usepackage{natbib}

\def\Msolar{$M_{\odot}$}

\newcommand{\Mo}{\rm{M}_{\odot}}

\usepackage{booktabs}
\usepackage{makecell}

\newcommand*\rfrac[2]{{}^{#1}\!/_{#2}}

\definecolor{pinegreen}{RGB}{1, 121, 111} 
\definecolor{violet}{RGB}{214,39,40}
\usepackage[breaklinks=true,colorlinks,urlcolor  = blue, linkcolor=violet,citecolor=pinegreen]{hyperref}  

\begin{document} 

\renewcommand{\arraystretch}{1.5} 

   \title{Looks can be deceiving}

  \subtitle{underestimating the age of single white dwarfs due to binary mergers}

   \author{K.D. Temmink\inst{1,2}
     \fnmsep\thanks{e-mail: \href{mailto:Karel.Temmink@ru.nl}{Karel.Temmink@ru.nl} }
          \and
          S. Toonen\inst{1,3}
          \and
          E. Zapartas\inst{1,4}
          \and
          S. Justham\inst{7,8,1,6}
          \and
          B.T. G\"ansicke\inst{5}
          }

   \institute{ Anton Pannekoek Institute for Astronomy, University of Amsterdam, 1090 GE Amsterdam,
The Netherlands 
         \and
             Department of Astrophysics/IMAPP, Radboud University Nijmegen, P.O. Box 9010, 6500 GL Nijmegen, The Netherlands
        \and     
            Institute for Gravitational Wave Astronomy and School of Physics and Astronomy, University of Birmingham, Birmingham, B15 2TT, United Kingdom
        \and 
            Geneva Observatory, University of Geneva, CH-1290 Sauverny, Switzerland
        \and
            Department of Physics, University of Warwick, Coventry CV4 7AL, UK
        \and 
            GRAPPA, University of Amsterdam, Science Park 904, 1098 XH Amsterdam, The Netherlands
        \and
            School of Astronomy \& Space Science, University of the Chinese Academy of Sciences, Beijing 100012, China
        \and 
            National Astronomical Observatories, Chinese Academy of Sciences, Beijing 100012, China
             }

   \date{Received October 10, 2019; accepted February 24, 2020}

  \abstract
   {White dwarfs (WDs) are important and abundant tools to study the structure and evolution of the Galactic environment. However, the multiplicity of WD progenitors is generally neglected. Specifically, a merger in a binary system can lead to a single WD, which could result in wrongly inferred quantities if only single stellar evolution (SSE) is considered. These mergers are linked to transients such as luminous red novae and Type Ia supernovae.}
   {We investigate the impact of binary evolution (BE) upon observable single WDs, and compare their properties to WDs formed through SSE. We assess the evolutionary channels and the age and mass distributions of the resulting single WDs.}
   {We employed \texttt{SeBa} to model the evolution of single star and binary populations. We synthesised the observable single WD population within $100$ pc, including cooling and observational selection effects. Additionally, we constructed models with different evolution and primordial population properties to study the effects on the properties of the resulting single WDs. }
  {White dwarfs from binary mergers make up about $10-30\%$ of all observable single WDs and $30-50\%$ of massive WDs.  On average, individual WDs take 
  $3.1-5$ times longer to form through BE than SE, and so appear ${\sim} 1$ Gyr younger than they are if BE is ignored. In all models, the effect of mergers on the age distribution is clearly noticeable. The median age typically increases by $85-430$ Myr and $200-390$ Myr for massive WDs. 
   Although abundant, we do not find evidence that WDs from mergers significantly alter the shape of the WD mass distribution. 
  }
  {Assuming SSE for inferring properties of single WDs gives rise to intrinsic errors as single WDs can also be formed following a binary merger. Strategies for mitigating the effect of mergers on the WD age distributions are discussed.
  }


   \maketitle
%

\section{Introduction}
\label{sec:intro}

White dwarfs (WDs) are the most common stellar remnants in the Milky Way (MW) and these objects account for the final evolutionary state of 97\% of stars \citep[e.g.][]{Fon01}. They are linked to several classes of transients, such as Type Ia  supernovae \citep[e.g.][]{Whe73, Ibe84, Web84, Mao14,Liv18}  and through their abundance, WDs are powerful tools to study the structure, formation, and evolution of the Galactic environment. For example, WDs can be used  as stellar age indicators and to infer the star formation history \citep[see e.g.][]{Tre14,Kil19} in a relatively simple way. Typically, the age is estimated by the sum of the cooling time (from measuring the effective temperature combined with theoretical cooling tracks) and the pre-WD stellar evolution time (from measuring the WD mass combined with models of single stellar evolution [SSE]). 

However, the majority of stars are found in multiple systems \citep[see e.g.][for a review]{Moe17}. 
Stars in close binary systems can interact with one another, eventually leading to a merger and subsequently a single WD. These single WDs formed through binary mergers can have vastly different evolutionary histories than those formed by isolated single stars. This implies that considerable inaccuracies could arise in studies that rely on models of SSE to infer properties of WDs (such as their ages) and derived quantities (such as the star formation history and the evolution of structure in the galaxy).

Merger remnants can be identified in different ways. One example is their possible rapid rotation due to the additional angular momentum of the pre-merger binary orbit. However, it is uncertain if the enhanced rotation would persist to the WD stage, for example if the merger remnant experiences significant wind mass loss before WD formation, such as for blue stragglers. 
An abnormal rotation profile or stellar structure may also be identified using asteroseismology \citep[e.g.][]{Com19, Bec12, Cor18}. 

Unfortunately, to the best of our knowledge, no star has been identified yet as a merger remnant using~asteroseismological techniques. As such, many merger remnants may be hiding between the single WDs. 

Stellar mergers are common outcomes of binary evolution (BE). As an example we mention the high rate of transients that are linked to stellar mergers, such as luminous red novae \citep{Koc14}. 
In the context of WDs, some claims are based on the mass distribution and the alleged excess of WDs with masses around $ 0.8 M_\odot$ (but see Section \ref{sec:massive_wds}). Lastly, we note that the binary fraction of a typical WD progenitor (A-type stars) is ${\sim}45\%$ \citep{derosa2014},  whilst the observed WD binary fraction is ${\sim}25\%$ \citep{Hol08, Too17, Hol18}. 
\cite{Too17} show that this tension is largely removed when stellar mergers are taken into account. From a theoretical perspective, these authors find that about $10-30\%$ of single WDs are expected to have a binary origin. 

In this work, we systematically study the effect of binary mergers upon the population of single WDs and their properties following a population synthesis approach. We describe our modus operandi in Section \ref{sec:method}. 
In Section \ref{sec:results}, we present results pertaining to the initial-final mass relation (IFMR) of single WDs (Section \ref{sec:ifmr}), their masses (Section\ref{sec:mass}) and ages (Section\ref{sec:age}). We close with our findings for the merger rates in Section \ref{sec:rates}. 
In Section \ref{sec:disc} we critically review our results and discuss the implications of stellar mergers for the use of WDs as age indicators. We end with a summary of our most important results in Section \ref{sec:concl}.


\section{Method}
\label{sec:method}
\subsection{\texttt{SeBa} \;-- A fast stellar and binary evolution code}\label{sec:seba}
In this work, we employ the binary population synthesis (BPS) code \texttt{SeBa}\footnote{The \texttt{SeBa} code is incorporated into the Astrophysics MUlti-purpose Software Environment, or AMUSE, and can be downloaded freely at \href{http://amusecode.org/}{amusecode.org}.} \citep{Por96, Too12} to simulate large numbers of single stars and binary systems. Using \texttt{SeBa}, we evolve a population of isolated single stars from the zero-zge main sequence (ZAMS) until the remnant phase. Furthermore, we use \texttt{SeBa} to generate large populations of binary systems on the ZAMS, simulate their subsequent evolution and extract those systems that merge and whose merger remnants form a WD - directly after the merger or with a delay.  
At every time-step, processes such as stellar winds, mass transfer, angular momentum loss, common envelopes (CEs), gravitational radiation and stellar mergers are considered with the appropriate prescriptions. 

The main cause for discrepancies between different BPS codes is found in the choice of input physics and initial conditions \citep{Too14}. In order to evaluate systematic uncertainties in our predictions for the binary populations, we construct a total of six BPS models; a fiducial model, and six additional models that differ from our default model in a single aspect. We employ four models with different assumptions regarding important phases in BE; that is CE evolution and the merger phase in Section\,\ref{sec:ce}~and~\ref{sec:mergers} respectively. We use a fifth model that differs with respect to the primordial binary population (Section\,\ref{sec:init}). A summary of our models can be found in Table \ref{table:summary_nrs}.

\subsection{Primordial population}
\label{sec:init}

We generate a population of single stars using a Monte Carlo approach. The metallicity is assumed to be solar, and the initial masses are drawn according to a Kroupa initial mass function \citep[IMF;][]{Kro93} in the range $0.95 \leq M \leq 10 M_\odot$. Our binary populations are initialised with a similar approach, whilst assuming solar metallicities and a Kroupa IMF for the masses of the initially more massive stars in each system. Throughout this work, we refer to these initially more massive stars as the primary stars of their respective binary systems. Properties related to primary stars are denoted with a subscript '1', whilst properties of the initially less massive companion stars are denoted with a subscript '2'. 
For the normalisation of our simulations, we consider primary masses in the range $0.1-100M_\odot$. For our fiducial model, we initialise the binary population according to the following distribution functions:
\begin{itemize}
\item Initial secondary masses $M_2$ are drawn from a uniform mass-ratio distribution with $0 < q\equiv \rfrac{M_2}{M_1} \leq 1$ \citep{Rag10, Duc13}. 
\item The semi-major axes are drawn from a distribution flat in $\log a$ \citep{Abt83}. The initial orbits are initialised with an upper limit of $ 10^6 R_\odot$, and a lower limit that ensures that neither component fills its Roche lobe on the ZAMS.
\item The orbital eccentricities are drawn from a thermal distribution between $0$ and $1$ \citep{Heg75}.
\item We assume a constant binary fraction of $50\%$ \citep[but see e.g.][]{Duc13, Moe17}.
\end{itemize}
In addition to our default model, we employ model "DM91" in which the orbits are initialised with periods drawn from a log-normal distribution \citep{Duq91} with a mean $\log(P/\text{days})$ of $4.8$ and a standard deviation of $2.3$.

\subsection{Mass transfer}
\label{sec:ce}

 A detailed overview of the treatment of various aspects of BE in \texttt{SeBa} can be found in \citet{Por96}, \citet{Nel01}, \citet{Too12},and \citet{Too13}. In the following, we provide a brief overview of those that are the most relevant to this work.
When one or both of the binary components fill their Roche
lobes, matter can flow from the donor star to the companion star. 
Whether mass transfer is stable depends on the reaction to the change in mass of the stellar radii and Roche lobes \citep[$\zeta \equiv \frac{dlnR}{dlnM}$; see e.g.][]{Hje87,Ge10,Ge15}. In \texttt{SeBa}, the stability of mass transfer is determined by comparing $\zeta$ of the donor star and related Roche lobe. The values of $\zeta$ for the donor star are based on detailed considerations \citep[see Tbl. A.1 in ][]{Too12}. The \texttt{SeBa} tool employs different values of $\zeta$ for the various phases in stellar evolution.
The efficiency of mass accretion depends on the stellar type of the accretor.
For normal, hydrogen-rich stars, hydrogen is accreted to the envelope, bounded by the thermal timescale of the accretor star, multiplied by a factor that depends on its effective radius and Roche lobe radius (for more details, see appendix A.3 in \cite{Too12}). Alternatively, accretion of helium-rich material is limited by the Eddington limit. Helium-rich stars are treated in a similar manner, where helium takes the place of hydrogen. Material not accreted is assumed to leave the system with $2.5$ times the specific angular momentum of the binary. This value was calibrated by previous studies where \texttt{SeBa} was employed to study the formation of high-mass X-ray binaries \citep{Por96,Nel01b}. For accretion onto WDs, model 'NSKH07' from \cite{Bou13} is used. This model is based on \cite{Nom07} for hydrogen-rich accretion, including updates from \cite{Hac08}, without wind stripping. For helium-rich accretion, this model is based on \cite{Kat99} and\cite{Hac99}. Any excess material is assumed to leave the system with the specific angular momentum of the accretor. 

If the mass transfer is unstable, the rate of mass transfer persistently increases in reaction to the mass loss of the donor star and a runaway situation ensues.
A CE is formed that engulfs both stars. This CE is not co-rotating with the binary, and as a consequence of drag forces inside the CE, the binary components spiral in to tighter orbits, which heats up the CE and transfers angular momentum. The CE phase ends when either the envelope is ejected from the system, leaving behind a more compact binary, or when the two components merge, leaving behind a single merger remnant. 
If a CE is survived by a binary, severe orbital shrinkage and the loss of the donor envelope can dramatically alter the subsequent evolution of the system. For an comprehensive review of CE evolution, see \cite{Iva13}. 

This short-lived phase in BE plays an important role in the formation of compact binaries, such as the progenitors of Type Ia supernovae, X-ray binaries and double neutron stars. Despite the importance of the CE phase and great efforts of the community in terms of both analytic and computational treatments, in addition to observational endeavours, this phase remains poorly understood. 
Thus, in BPS codes, simplified prescriptions are often adopted to model the CE evolution. The canonical model for the CE phase is the so-called $\alpha$-prescription \citep{Pac76,Web84,liv88,DeK87,DeK90}, which is based on the energy budget of the binary system. The energy to unbind the envelope comes from the release of orbital energy ($\Delta E_{\rm orbit}$) due to the shrinkage of the binary orbit. This process might not be perfectly efficient and a parameter $\alpha$ is introduced to quantify the fraction of released orbital energy consumed to unbind the CE as follows:
\begin{align}
E_{\rm bind} = \frac{G M_{\rm d} M_{\rm e}}{\lambda R} = \alpha \Delta E_{\rm orbit}
\label{eq:ebind}
\end{align}
where $E_{\rm bind}$ is the binding energy of the envelope, $M_{\rm d}$ is the total mass of the donor star, $M_{\rm e}$ is the mass of its envelope, and $R$ is its radius. The parameter $\lambda$ is dimensionless and depends on the structure of the donor star. The change in orbital energy is calculated as:
\begin{align}
\Delta E_{\rm orbit} = \frac{GM_{\rm a}}{2} \left(\frac{M_{\rm c}}{a_{\rm f}} - \frac{M_{\rm d}}{a_{\rm i}}  \right)
\label{eq:eorbit}
\end{align}
where $M_{\rm c}$ is the mass of the core of the donor star, $M_{\rm a}$ is the mass of the companion star and $a_{\rm i}$ and $a_{\rm f}$ are the semi-major axis pre- and post-CE, respectively. When solving for the post-CE orbital separation, the parameters $\alpha$ and $\lambda$ naturally appear as a product. Therefore, they are often treated as a single parameter $\alpha\lambda$. If the envelope cannot be ejected by use of the orbital energy or if either of the two components fills their new Roche lobes after the CE, we assume the system merges.

An alternative model for classically unstable mass transfer is the $\gamma$-prescription \citep{Nel00}, in which the angular momentum budget is considered. This prescription is motivated by the observed distribution of mass ratios of double WD (DWD) systems that cannot be explained by the $\alpha$-prescription. In this prescription, angular momentum from the spiralling-in is used to unbind the CE. The fraction of the angular momentum that is used towards unbinding the envelope is represented by the $\gamma$ parameter, that is:
\begin{align}
\frac{\Delta J}{J_{\rm init}} = \frac{J_{\rm init} - J_{\rm fin}}{J_{\rm init}} = \gamma \frac{M_{\rm e}}{M_{\rm d} + M_{\rm a}}
\end{align}
where $J_{\rm init}$ and $J_{\rm fin}$ are the angular momentum of the pre- and post-CE binary, respectively.

Uncertainty in the outcome of a CE phase can strongly affect the properties of binary populations and the evolution of individual systems \citep[e.g.][]{Han02, Ruit07, Too12}. To investigate in what manner our results are dependent upon assumptions regarding the modelling of the CE phase, we employ several different models \citep{Too17}.

In our fiducial model, we use the $\alpha$-prescription to model the CE with $\alpha\lambda = 2$ \citep{Nel00,Nel01b}, which is based on the reconstruction of the second phase of mass transfer for observed DWDs. 

We furthermore employ two models; $\alpha$-ineff and $\alpha$-eff. In the former model, we assume a CE that uses $\Delta E_{\rm orbit}$ less efficiently with $\alpha\lambda = 0.25$, motivated by studies from  \cite{Zor10,Zor14,Too13,Cam14}, who deduced a less efficient CE phase from the population of observed post-CE binaries. In the latter model, we study the effect of a CE that is more efficient, where $\alpha\lambda = 5$. Lastly, we employ an additional model $\alpha\gamma$, where we employ the $\gamma$-prescription with $\gamma = 1.75$. If, however, the companion is a degenerate object, or the CE is the result of a Darwin-Riemann instability, we employ the $\alpha$-prescription with $\alpha\lambda=2$ instead \citep{Nel01}. If both stars are on the MS, we assume that unstable mass transfer leads to a merger.

\subsection{Stellar mergers}
\label{sec:mergers}

In \texttt{SeBa}, we simulate the merger of a binary in the following way. The properties of the merger remnant depend upon the type and masses of the binary components and are described below for each combination of progenitor stellar types. Most relevant in this study is the post-merger mass of the star (i.e. how much mass is lost from the system during the merger), and its evolutionary age.  
Unless stated otherwise, we assume that the merger remnant is of the same stellar type (i.e. the same evolutionary phase) as the donor star (the star that fills its Roche lobe) of the progenitor system. The merger remnant can both appear younger (rejuvenation) or older compared to the donor star, which is modelled as described in appendix A.2 of \cite{Too12}. 

The mass of the merger remnant is determined as follows:\\
If both component stars are on the main sequence (MS), we assume that the two stars merge fully conservatively in terms of mass, resulting in a more massive MS merger remnant. This is in agreement with results from \cite{Gleb13}, who find that $\lesssim 10 \%$ of the total mass is lost in similar mergers. These types of mergers would appear in the Hertzsprung-Russell diagram as so-called blue stragglers \citep[e.g.][]{Zin76}. \\
In systems where the donor star is a giant (post-MS) and its companion star is a MS star, the mass of the companion is added to the envelope of the giant fully conservatively. \\
If both components are post-MS stars, their cores merge without loss of mass. Regarding the envelope of the merger remnant, we assume half of the hydrogen-rich envelope of the companion star is lost from the system \citep[but see e.g.][]{Mac17,Kam18,Mac18}. If the donor star is a giant star, and its companion is a WD, the mass of the WD is added to the core of the donor star fully conservatively. \\
If a MS star merges with a WD, we assume a giant star is formed. The WD forms the new core of the merger remnant without loss of mass. If the mass of the accreted WD is large enough for helium ignition to occur, the remnant evolves along the horizontal branch (HB) as a core helium burning star. If the mass of the companion is below the helium ignition limit, the remnant is a star evolving along the red giant branch (RGB). \\ 
Lastly, if both progenitor components are WDs, and the combined mass is below 1.38\Msolar, the remnant is also a WD with a mass that is equal to the sum of the masses of the progenitors. This is in agreement with hydrodynamic simulations of double WD mergers, which find that typically ${\sim}10^{-3}M_{\odot}$  of mass ejected from the system \citep[e.g.][]{Lor09}. If the combined mass is above 1.38\Msolar, we assume that the merger leads to a disruption of the remnant, and hence no WD is left behind.
Below 1.38\Msolar, we make the simplifying assumption that there is no (re-)ignition in the merger remnant. Thus we ignore the evolutionary phase of R Coronae Borealis variables (see also Section \ref{sec:limitations}) and the possibility of sub-Chandrasekhar explosions during double WD mergers \citep[e.g.][]{Web84,Van10b}. Regarding the latter, we note that the far majority of these systems have a combined mass above 1.38\Msolar in our simulations \citep[see also][]{She18}.\\
We note that in this framework we consider naked helium stars as evolved stars. All material is accreted to their envelopes conservatively. 

For mergers that occur through a CE event, we define an additional model "$\alpha$-noncon". In this model, we allow for partial envelope ejection during a CE merger. This partial envelope loss is in addition to any mass lost during the merger event itself and can therefore lead to different merger remnant evolution. Making the substitutions $M_e \rightarrow \Delta M, M_c \rightarrow M_d - \Delta M$, the amount of mass lost from the envelope of the donor $\Delta M$ can be calculated from equations \ref{eq:ebind} and \ref{eq:eorbit}:
\begin{align}
\left( \frac{M_d}{\alpha\lambda R} + \frac{M_a}{2 a_{\rm RLOF}} \right) \cdot \Delta M = \left(\frac{M_a M_d}{2 a_{\rm RLOF}} - \frac{M_a M_d}{2 a_i}\right)
\label{eq:noncon}
\end{align}
where $a_i$ is the pre-CE orbital separation as before, and $a_{\rm RLOF}$ is the widest orbit at which either the companion star or the core of the donor star fills its Roche lobe.

\begin{table}
\caption{ Description of the BPS models }
\label{table:summary_nrs}      
\centering          
	\begin{tabular}{@{}l|l@{}}
		\toprule
		Model name              & Model description                                   \\ 
		\midrule
		default                 & $\alpha\lambda=2$                                          \\
		$\alpha$-noncon & non-conservative CE during merger                                \\
		$\alpha$-ineff         & $\alpha\lambda=0.25$                                \\
		$\alpha$-eff         & $\alpha\lambda=5$                                  \\
		$\alpha\gamma$          & $\gamma=1.75$, $\alpha\lambda=2$          \\
		DM91                     & log-normal distribution of initial periods                                      \\
		\bottomrule
	\end{tabular}
\end{table}

\subsection{Modelling the Milky Way}
\label{sec:mw}
To allow for direct comparison to observations, we construct model samples of the WD population within $100$pc of the Sun, which we refer to here as our `model Milky Ways (MWs)'. We adopt a stellar formation history and take into account WD cooling and observational selection effects as follows.

The construction of the model MWs is based upon the "cSFR" model from \cite{Too17}. We assume a Galactic age of $10$Gyr \citep[see e.g.][]{Leg98,Osw96,Kil17} and we employ a simple stellar formation history with a constant star formation rate \citep[e.g.][]{Woo92} of $3M_\odot$yr$^{-1}$. We constrain ourselves to physical distances $d \leq 100$pc, and assign random positions to our synthetic WDs. We assume that within this volume stellar systems are distributed uniformly.

To model the cooling of the WDs, we use WD cooling tracks from {\cite{Hol06,Kow06,Tre11,Ber11}\footnote{See also the official website for these cooling tracks: \href{http://www.astro.umontreal.ca/~bergeron/CoolingModels/}{http://www.astro.umontreal.ca/~bergeron/CoolingModels/} } }. From these cooling tracks and the synthetic positions of the WDs, we calculate their \textit{ugriz}-band magnitudes. Finally, we use relationships by \citep{jordi10} to convert our \textit{ugriz} values to the magnitude in the Gaia $G$ band. Then, we select only those systems for which $G \leq 20$, such that we only model WDs that are observable by Gaia \footnote{As indicated on the official Gaia website: \href{https://www.cosmos.esa.int/web/gaia/science-performance}{https://www.cosmos.esa.int/web/gaia/science-performance} }.

\section{Results}
\label{sec:results}

\begin{figure}[h!]
	\centering
	\includegraphics[width=\hsize]{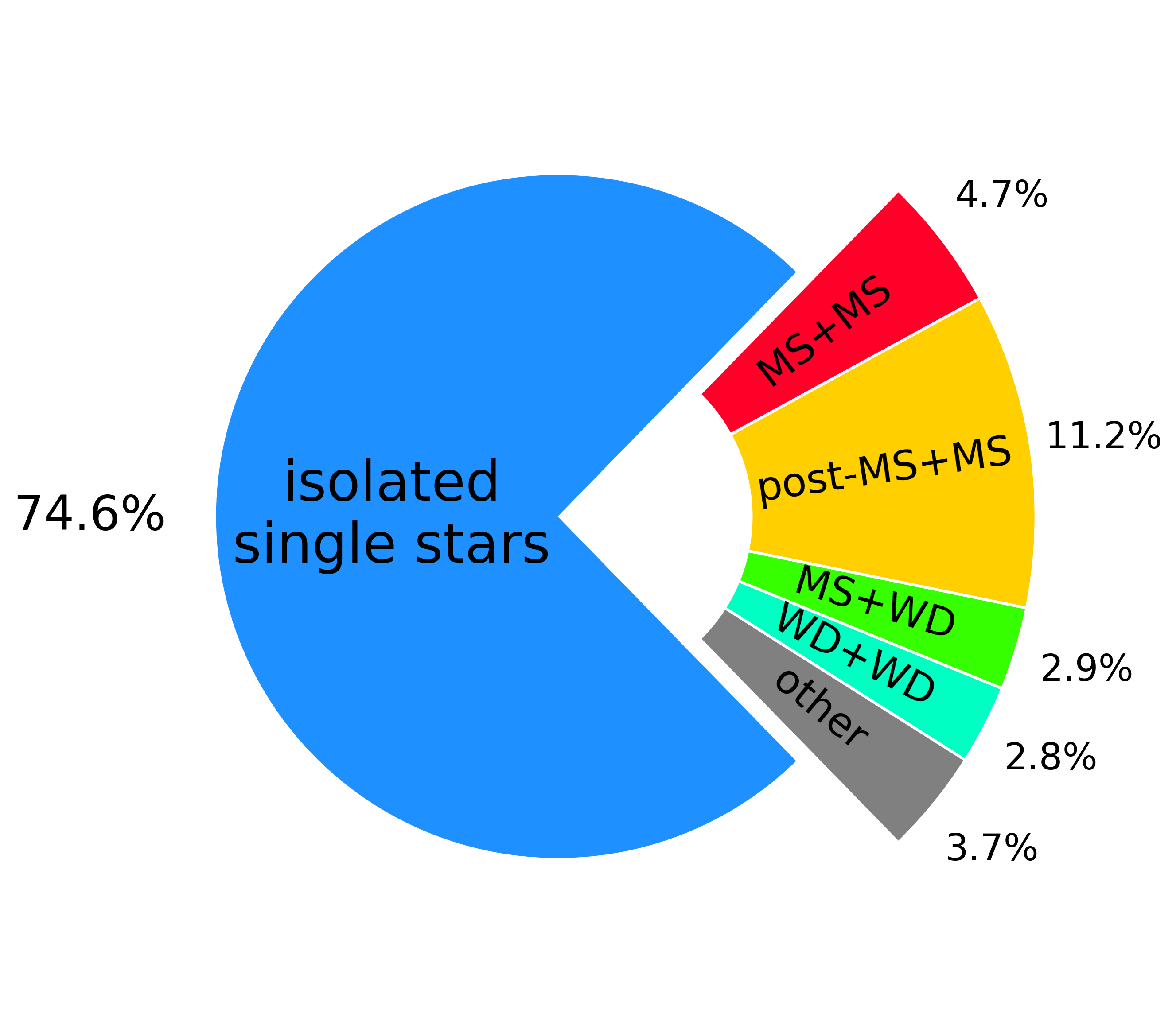}
	\caption{Origin of WDs in our default model: isolated single stars (blue) and binary mergers (other colours). For binary mergers, several important types of merger progenitors are highlighted.
	}
	\label{fig:sin_bin_WDs_default}
\end{figure}

From our simulation of isolated single stars and binary stars, we find that the contribution of binary mergers to the  total number of currently observable single WDs in the $100$pc sample is about ${\sim} 10-30 \%$ (Tbl. \ref{table:results_table})\footnote{In this study we do not consider WDs in (un-)resolved binaries. Taking these systems into account as in \cite{Too17} would lower the fraction of single WDs formed through binary mergers by roughly $5\%$.} 
This is in agreement with results from \cite{Too17} from a similar analysis of the $20$pc WD sample. 
In our default model, binary mergers contribute about $25\%$ to the total number of single WDs. They form from a variety of merger channels, amongst others MS+MS, `post-MS'+MS, and WD+WD mergers (Figure\,\ref{fig:sin_bin_WDs_default} and Section\,\ref{sec:ev}). 
The fraction is not sensitive to whether the CE phase is conservative in terms of mass. However, the efficiency at which orbital energy is used during a CE phase affects the fraction significantly because it is directly related to the amount of orbital shrinkage during the CE. 
In our model $\alpha$-ineff, the CE phase proceeds less efficiently and the orbit shrinks relatively more. Hence, the number of mergers leading to a single WD increases by about $30\%$ compared to our default model. In contrast, a more efficient CE phase with less orbital shrinkage leads to about $20\%$ less mergers. Similarly, adopting the $\gamma$-CE prescription results in $15\%$ less single WDs from mergers.
The properties of the primordial binary population also have an effect on the number of single WDs produced. If the primordial periods are drawn from a log-normal distribution as in model DM91, the total number of single WDs decreases significantly, forming slightly less than half the amount of single WDs as our default model. This is because this model encompasses fewer primordial binaries with short periods, hence fewer binaries interact and merge.
A comprehensive summary of our findings, is provided in Table \ref{table:results_table}.

\subsection{Initial-final mass relation}
\label{sec:ifmr}

\begin{figure}[h!]
	\centering
    \includegraphics[width=\hsize]{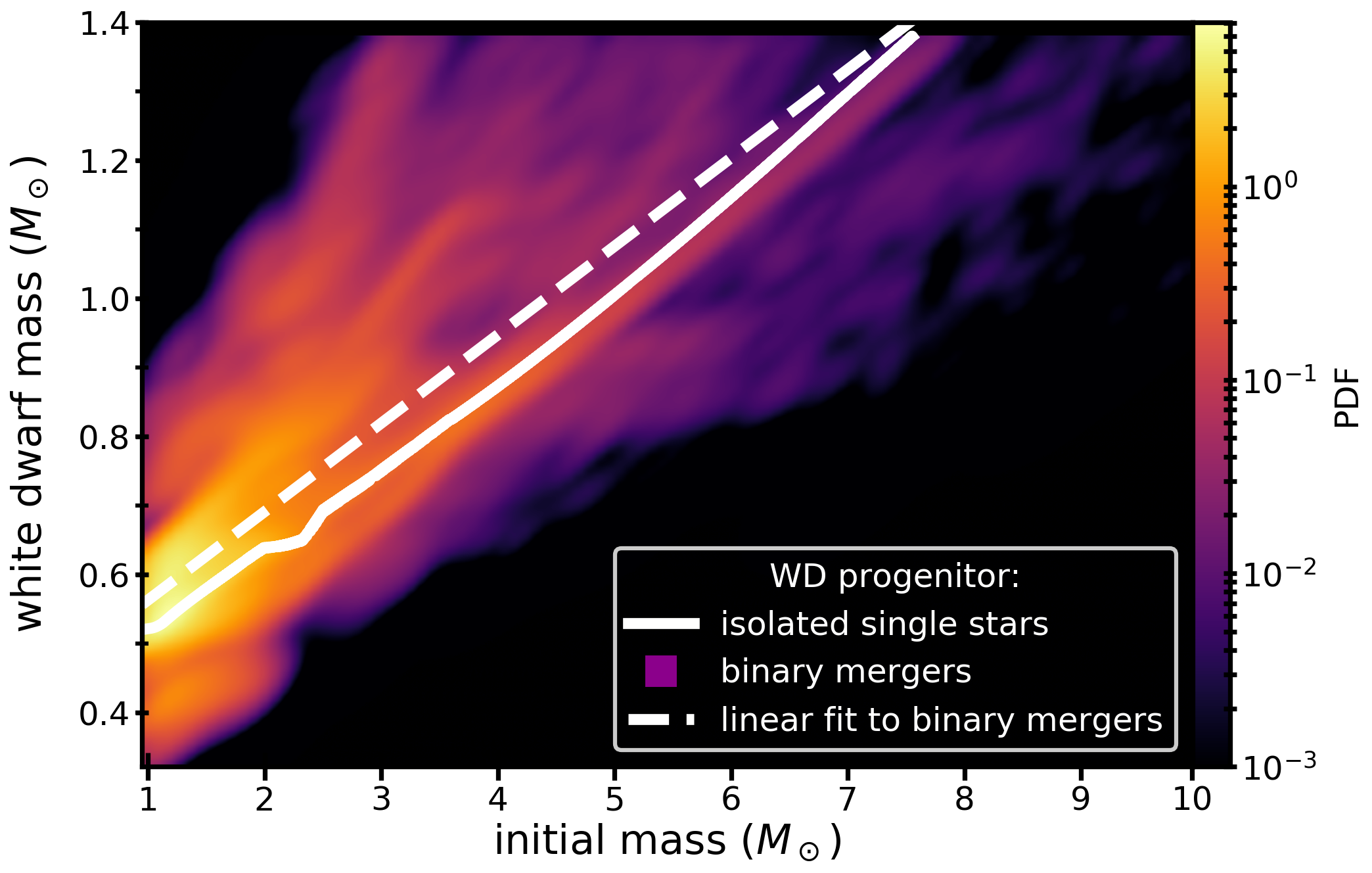}
	\caption{Initial-final mass relations (IFMRs) for single WDs. The solid white line indicates the IFMR for stars that evolve as isolated single stars. The colours represent the IFMR for single WDs that form as a result of a merger event in a binary at some point during the evolution of the progenitor system. More precisely, the colours logarithmically indicate the 2D probability density function (PDF) obtained from our data through the kernel density estimation method. We plot the initial mass of the initially more massive star against the mass of the single WD formed after the merger event. In this figure, we assume all stars are formed simultaneously $10$Gyr ago, and do not include observational selection effects. The dashed white line is a linear fit ($M_{\rm WD} = 0.128M_{\rm init} + 0.435$) to the binary merger IFMR, obtained through linear regression of all ($M_{\rm WD},M_{\rm init}$) points in our data.}
	\label{fig:IFMR_comp}
\end{figure}

In Figure \ref{fig:IFMR_comp}, we present the IFMR for WDs formed through SSE as well as binary mergers. We find that the IFMR for binary mergers is not a one-to-one relation, as is the case for isolated single stars. Mass transfer in binary stars can both increase and decrease the final mass of the single WD for a given initial stellar mass. Additionally, the converse is also true: there is a relatively large spread of initial masses corresponding to the same final single WD mass. On average, however, WDs formed through binary mergers are more massive  than those formed by isolated single stars, as indicated by the linear fit to the IFMR from binary mergers in Figure \ref{fig:IFMR_comp}.

Furthermore, binary interactions can lead to the formation of lower-mass WDs than can be explained by standard SSE only \citep[here $\lesssim 0.51M_\odot$, but see][]{Ibe85, Zen19}, by stripping one or both of the component stars in a mass transfer event. Additionally, WDs of a given mass can be produced by stars that are initially more massive than when only isolated single stars are considered. We note that this can also affect single WDs that are formed from binaries without mergers, for example when one component explodes in a Type Ia supernova, as studied in for example \cite{Jus09}.

\subsection{Evolutionary channels}

\label{sec:ev}
\begin{figure}[h!]
	\centering
    \includegraphics[width=\hsize]{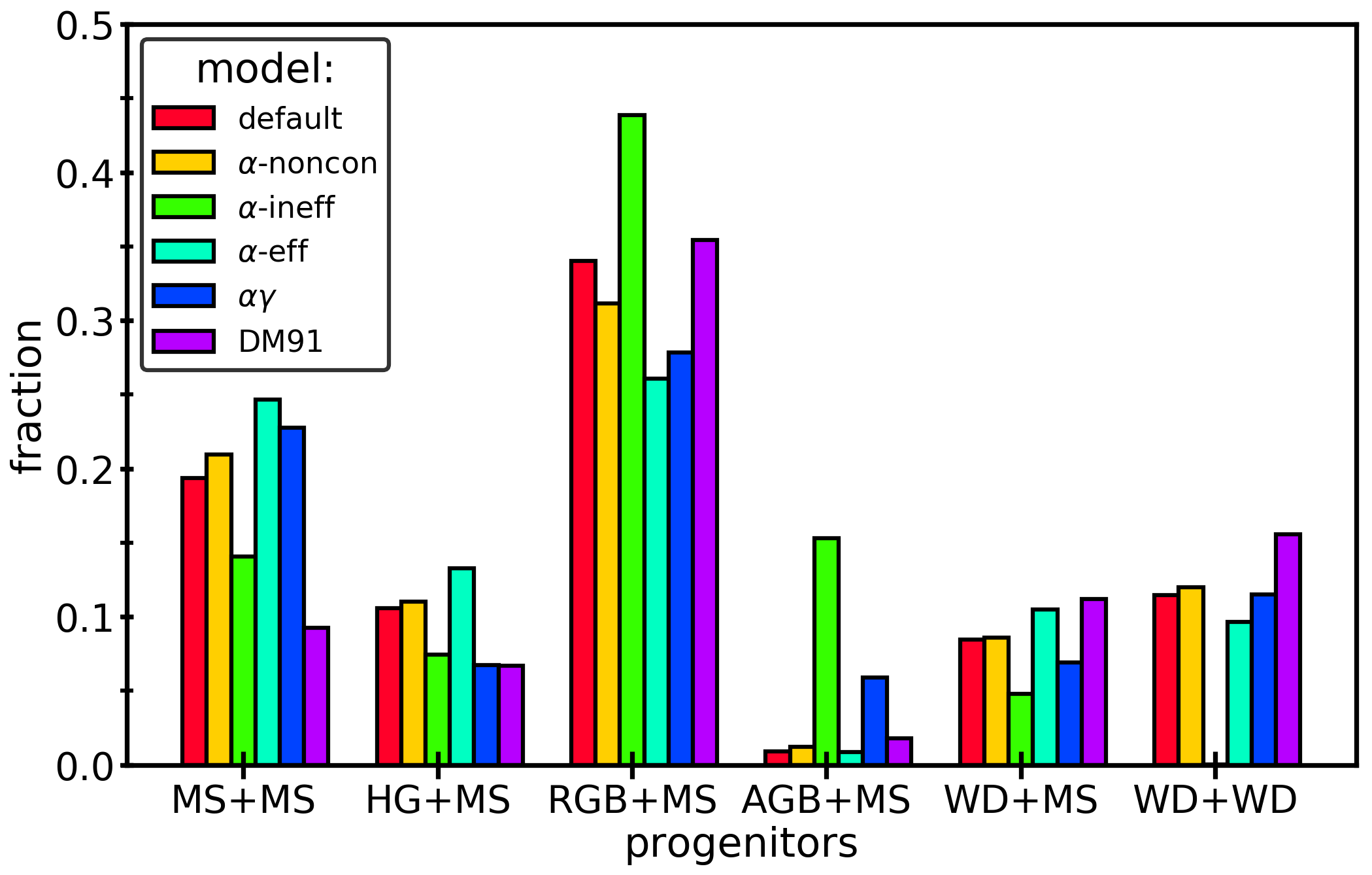}
	\caption{Fractional contributions of various formation channels to the total number of WDs from binary mergers. The colours of the bars indicate the results of different models. The height of the bars corresponds to the fraction of all single WDs that forms through a certain type of merger. The specific type of merger is indicated on the x-axis. For each type of merger, the colours correspond to the models in the same order as listed in the legend.
	}
	\label{fig:progenitor_contributions_all}
\end{figure}

There are several evolutionary pathways that can result in a binary system forming a single WD. The relative contribution of the most important evolutionary channels is shown in Figure\,\ref{fig:progenitor_contributions_all}. We find that the primary contribution to single WDs comes from mergers between a post-MS star and a MS star. In the default model this accounts for ${\sim} 45\%$ of the binary mergers (see Figure\,\ref{fig:sin_bin_WDs_default}). For our other BPS models, the contribution of mergers between post-MS stars and MS stars ranges from $\sim15\%$ to $\sim 65\%$. 

Typically, the second most important type of merger is a merger between two MS stars (possibly leading to a blue {straggler}), which contribute ${\sim} 15-25\%$ of the single WDs from binary mergers. In particular, this type of merger is more prominent in models in which there is less orbital shrinkage during the CE phase (that is models $\alpha$-ineff and $\alpha\gamma$), because the number of mergers that are \textit{not} of the MS+MS type is dependent on the value of $\alpha\lambda$.

The WD+WD mergers form $\lesssim 10-15 \%$ of all single WDs for all our models except for model $\alpha$-ineff, for which we find an insignificant number of double WD mergers. Mergers between stars on the asymptotic giant branch (AGB) and MS stars are only important for model $\alpha$-ineff and play almost no role in other models. 

In the following, we give a more detailed description of the initial parameter spaces and BE leading to the types of mergers discussed above.

\subsubsection{Channel 1: MS+MS mergers}	
The binaries in this channel have initially close orbits with semi-latera recta  {$a(1-e^2) \sim 2~-~20 R_\odot$}. The distribution of primary masses peaks around $1.5 M_\odot$ and falls off steeply to either side, which is a direct result of the combination of the IMF and the fact that the MS lifetime typically decreases with increasing stellar mass.

About $40\%$ of systems merge immediately at the onset of Roche-lobe overflow (RLOF). In about $60\%$ of systems there is a period of stable mass transfer preceding the merger. After several tens of Myr, the mass transfer becomes unstable and the system merges.

\subsubsection{Channel 2: HG+MS mergers}
In this channel, stars merge as the primary star makes the transition from the MS to the RGB, when stars occupy the Hertzsprung gap (HG) region in the HR diagram. 

Whilst the general shape of the primary mass distribution is similar, the typical HG+MS merger progenitor is slightly more massive than than MS+MS merger progenitors (their distribution peaks near $M=1.75\,M_\odot$). The semi-major axes are generally smaller than those of the RGB+MS channel (see below), but larger than those of the MS+MS channel, with $a(1-e^2) \sim 3-30 R_\odot$. Once the primary star fills its Roche lobe, dynamically unstable mass transfer results in a CE, and the components merge. In about $80\%$ of our systems, a period of stable mass transfer precedes the CE phase, which occurs when the stellar type changes. These systems typically lose a significant fraction ($\sim 20-30\%$) of the total mass in the system before the CE occurs.

\subsubsection{Channel 3: RGB+MS mergers}
In this formation channel, the primordial binary systems have initial orbits peaking at {$a(1-e^2) \sim 30 R_\odot$}, but spanning a large range $\sim 5-300 R_\odot$. The initial distribution of primary masses is similar to what we find in the MS+MS merger channel, peaking at a marginally higher mass ($\sim 1.6\,M_\odot$), and with a slightly shallower slope on either side of the peak. %

In this channel, the first and only moment of RLOF occurs when the primary is either in the Hertzsprung gap (HG) or on the RGB. For $\sim 80\%$ of systems, this situation is unstable from the onset, immediately resulting in a CE phase. The remaining $\sim 20\%$ have a short phase (tenths of Myr) of stable mass transfer, before eventually a CE is formed. All systems in this channel merge through a CE phase, and leave behind a more massive RGB star, enriched with the hydrogen that made up the companion star.

\subsubsection{Channel 4: AGB+MS mergers}
Single WDs that form through the AGB+MS channel find their origins in the widest of orbits with $a(1-e^2) \sim 50-1,800 R_\odot$.
Binaries in this channel have initially higher primary star masses $\sim 2-8 M_\odot$ %
When the primary fills its Roche lobe, dynamically unstable mass transfer leads to a CE phase and a merger of the system. The result of the merger is a new AGB star. After a short period ($\sim$tenths of Myr), this newly formed AGB star expels its envelope and forms a WD. 

\subsubsection{Channel 5: WD+MS mergers}
Systems forming a single WD through this channel have initial masses %
distributed similarly as for the aforementioned channels. However, the distribution of initial primary masses only extends up to $\sim 6 M_\odot$.  The systems start from a small range of wide orbits $a(1-e^2)\sim 10-60 R_\odot$. Typically, the primary fills its Roche lobe on the RGB ($\sim 90\%$), or otherwise on the AGB. This first phase of RLOF always results in a CE phase, eventually forming a binary with a WD and a MS companion.  
When the secondary fills its Roche lobe, the components merge and form an evolved star; depending on the mass of the accreted WD, the merger remnant is either on the RGB or on the HB (see Section \ref{sec:mergers}).

\subsubsection{Channel 6: WD+WD mergers}
For these systems, the distribution of initial orbits is different from the other BPS models. The DWD mergers initialise according to a bi-modal distribution of $a(1-e^2)$, which peaks at $\sim 15 R_\odot$ and $\sim 500 R_\odot$. The primary masses span the full simulated range, but favour intermediate masses $\sim 2.5 M_\odot$. 
In this channel, one to four mass transfer phases can occur, which are described in more detail in \cite{Nel01} and \cite{Too12}.  
In short, the first mass transfer phase typically occurs when the primary star is on the RGB (initially closer orbits) or AGB (initially wider orbits) and the  companion star is still on its MS. In $\sim 40\%$ of binaries, this first phase of RLOF is one of stable mass transfer. In other cases, a CE is formed that is expelled from the system. In those systems that have initially wider orbits, two CE phases occur before the single WD is formed. The second CE occurs when the companion star has evolved off the MS, typically when it is either on the RGB or AGB.

\subsection{White dwarf masses}
\label{sec:mass}

\begin{figure}[h!]
	\centering
    \includegraphics[width=\hsize]{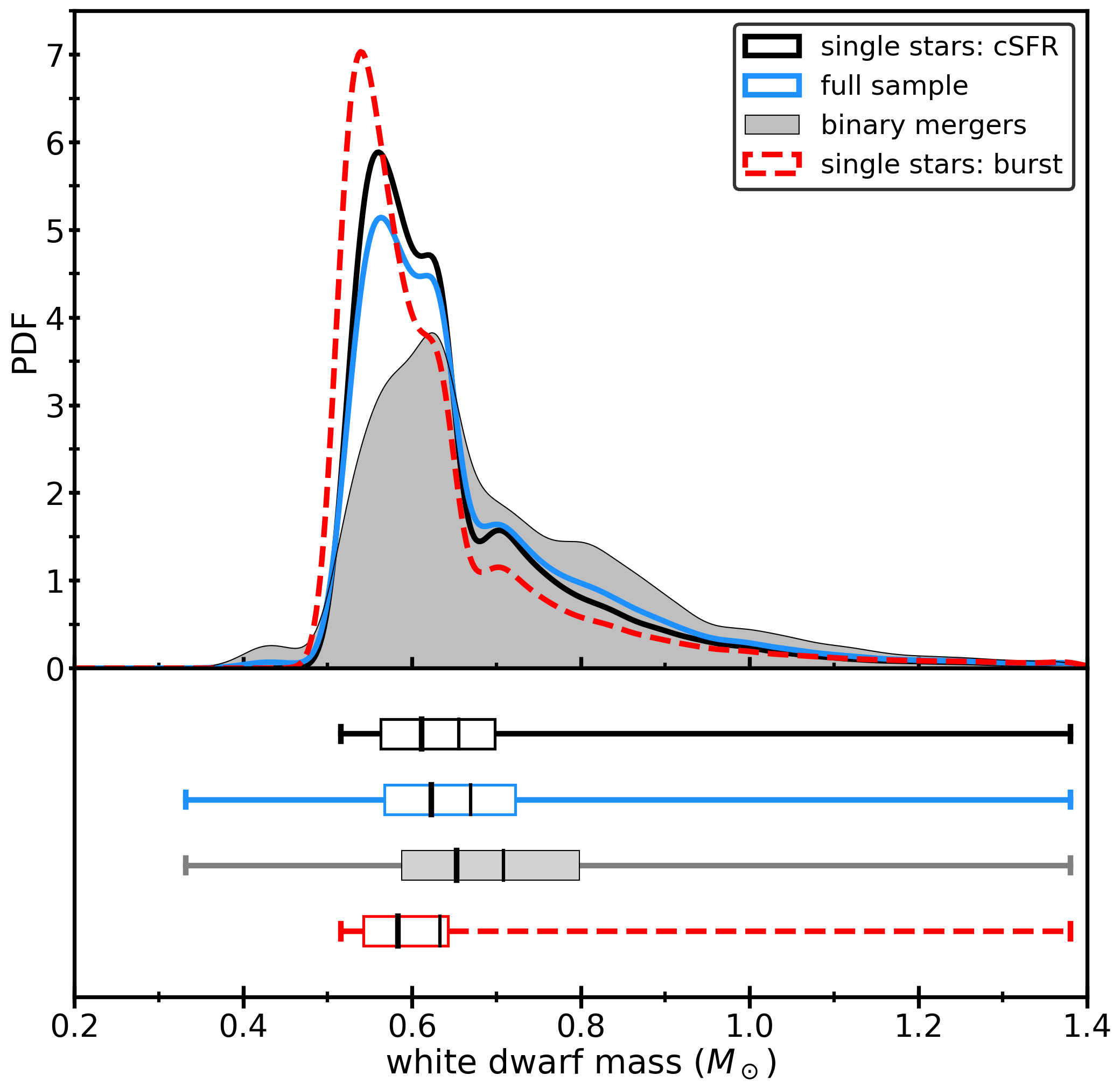}
	\caption{Mass distribution of single WDs. Top panel: The black solid line represents WDs within $100$pc formed by isolated single stars, including selection effects (see Section \ref{sec:mw}). The blue solid line indicates the distribution of masses in the full sample (binary merger and isolated single stellar progenitors). The shaded grey histogram represents the distribution of WD masses from binary mergers in our default BPS model.  Lastly, the dashed red line indicates the theoretical WD mass distribution resulting from a single burst of star formation $10$Gyr ago.
	In this figure, the PDFs are individually normalised to integrate to unity, and hence have different absolute scales. Bottom panel: Box plots of the aforementioned distributions. The whiskers span the full range of the distributions, and the width of the boxes indicates the inter-quartile ranges (IQRs). The thick (left) and thin (right) vertical lines indicate the median and mean values of the distribution respectively.}
	\label{fig:mass_distr_iss}
\end{figure}

\begin{figure}[h!]
	\centering
    \includegraphics[width=\hsize]{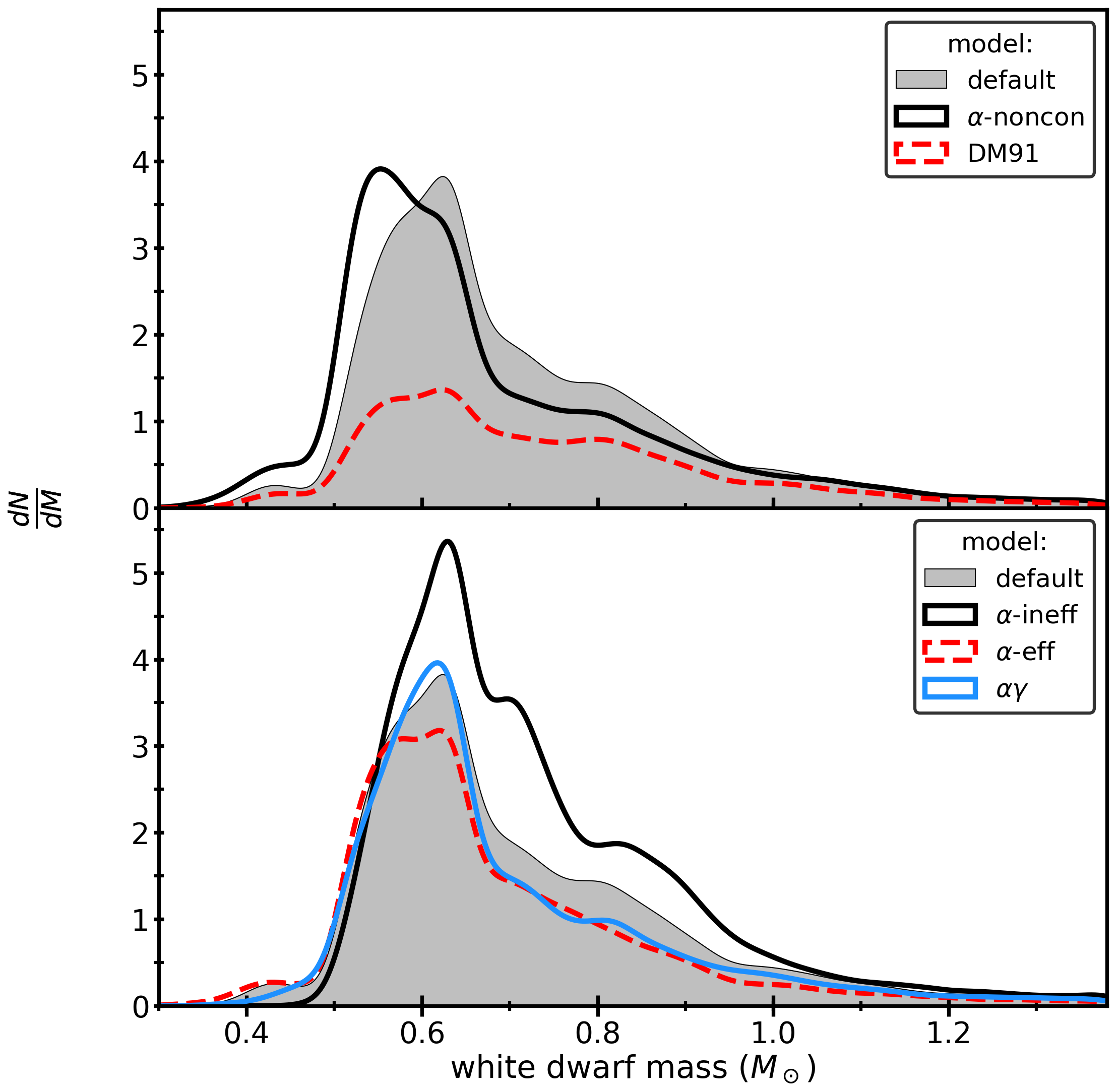}
	\caption{Mass distribution of WDs formed by binary mergers. In both panels, our default BPS model is shown in grey shading. The \textit{top panel} contains our results for our default model and its non-conservative CE counterpart, as well as for model DM91. The \textit{bottom panel} contains the results from our three different CE models. The PDFs shown in this figure have been re-scaled to represent the number of single WDs formed in each model relative to our default model, which is normalised to integrate to unity.
	}
	\label{fig:mass_distr_binmerge}
\end{figure}

The WD mass distribution formed by isolated single stars is given in Figure \ref{fig:mass_distr_iss}. The distribution of masses peaks at $0.59M_\odot$ with a median mass of $0.61M_\odot$ and a mean mass of $0.66M_\odot$, which is consistent with values inferred from observations \citep[e.g.][]{Lie05, Tre11, Gia12, Tre16}. 
There are minor features in the WD mass distribution around a WD mass of $0.61M_\odot$ and $ 0.7M_\odot$, which directly relates to the single stellar IFMR in \texttt{SeBa} (Sec.\,\ref{sec:ifmr}).

In Figure \ref{fig:mass_distr_iss} we also show the mass distribution of the single WDs from binary mergers in our default PS mode l. The distribution peaks at $0.64M_\odot$, with a median mass of $0.65M_\odot$ and a mean mass of $0.71M_\odot$. Overall, the distribution is broader than that from the single stellar population, because the merger remnants typically result in more massive WDs for the same ZAMS (primary) mass. 
The full sample includes WDs from both our isolated single stellar sample and our default binary merger sample at a initial binary fraction of $50\%$. This mass distribution resembles that of the single stellar population, but with a greater number of massive WDs.

In Figure \ref{fig:mass_distr_binmerge}, we present the mass distribution of WDs formed by binary mergers for all of our BPS models. We find that the mass distribution is generally not sensitive to the limiting distance nor the limiting Gaia G magnitude in all our models.The main (but small) effect is in the number of very massive WDs ($M \gtrsim 1.0 M_\odot$). The exact shape of the mass distribution is dependent on the model. In model $\alpha$-noncon, where we allow for partial envelope ejection during a CE, the distribution peaks at a lower mass (i.e. $0.53 M_\odot$) and there is a large increase of WDs with masses $\lesssim 0.6M_\odot$. This is a direct consequence of the extra mass lost during CE phases, since less mass can be added to the core of the merger remnant after the CE, resulting in more low-mass WDs. Changing the initial distribution of periods (model DM91) mostly affects the number of WDs formed, but also results in a slight increase in the fraction of WDs with masses ${\sim}0.8M_\odot$.  

The bottom panel of Figure \ref{fig:mass_distr_binmerge} shows the WD mass distribution for various assumptions regarding the CE phase.
If the CE phase uses orbital energy less efficiently (model $\alpha$-ineff), the fraction of WDs found at the low-mass end of the distribution is lower, whilst the fraction of WDs at $M \geq 0.8$ is higher. In this model, the WDs from binary mergers are typically formed by mergers involving evolved  primary stars (on the RGB and AGB). These stars generally have more massive cores, resulting in more massive WDs.

Conversely, modelling the CE as more efficient (model $\alpha$-eff) results in an excess of WDs with $M \lesssim 0.6 M_\odot$ when compared to our default model. These WDs typically form through the merger of a helium WD with a low-mass MS companion. Additionally this model results in the largest number of single low-mass ($M \leq 0.51\, M_\odot$) WDs of all models, which are the result of mergers of double helium WD systems (see also Section \ref{sec:limitations}).

In summary, the efficiency of the CE phase has a direct impact on the single WD mass distribution. If the CE proceeds more efficiently, less mergers occur during a CE phase. Additionally, equations \eqref{eq:ebind} and \eqref{eq:eorbit} imply that either the `typical' donor star is less evolved when it fills its Roche lobe or that the accretor star is of lower mass. We directly observe these effects as a tendency towards lower single WD masses for larger values of $\alpha\lambda$.

\subsection{White dwarf ages}
\label{sec:age}

\begin{figure}[h!]
	\centering
    \includegraphics[width=\hsize]{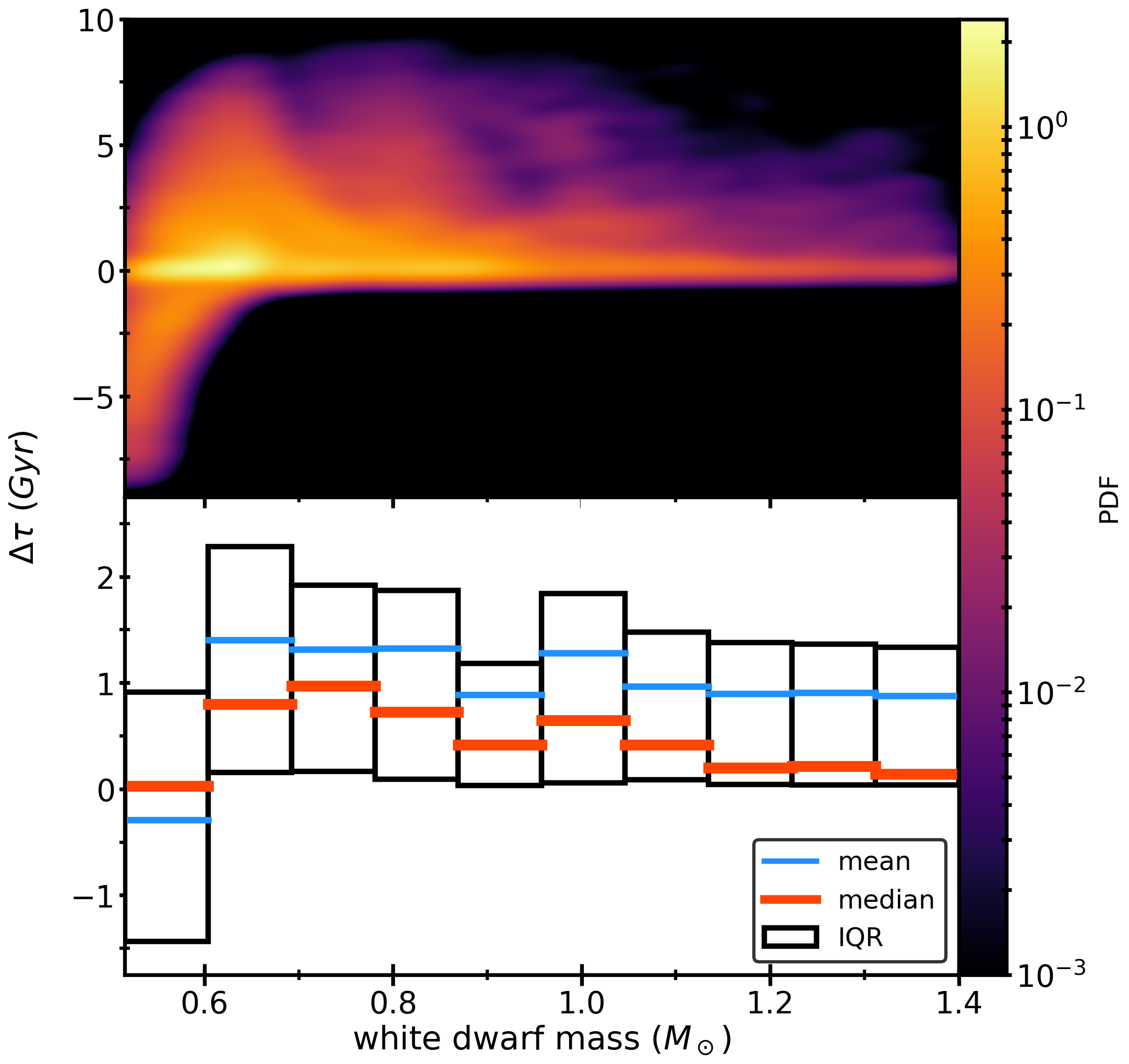}
	\caption{Differences in inferred WD formation time $\Delta\tau \equiv \tau_{\rm BE} - \tau_{\rm SSE}$ (see text) for WDs from binary mergers as a function of WD mass in our default model. Top: The colour logarithmically indicates the 2D PDF inferred from our data. Bottom: Box plots of the $\Delta\tau$ within indicated mass bins. The blue and red lines indicate the mean and median values of the distribution, respectively. The black boxes represent the IQR, which is equal to the difference between the upper and lower quartiles of the distributions.
	}
	\label{fig:dtau_mwd}
\end{figure}

\subsubsection{Formation times}
\label{sec:formation_times}
In our discussion of the WD formation times and ages, we do not consider WDs with masses less than $0.51 M_\odot$ in this work because they cannot be formed by SSE in $10$Gyr (as briefly discussed in Section \ref{sec:ifmr}). The number of WDs with masses below $0.51 M_\odot$ that we remove from our sample  is typically of the order of $5\%$ of the total number of single WDs formed by binary mergers.

For the WDs from binary mergers, we compare two timescales: The time it takes to form a single WD through BE starting at the ZAMS, $\tau_{\rm BE}$, and the formation time $\tau_{\rm SSE} (M_{\rm WD})$ we would derive for the same object if only SSE is considered. 
We estimate $\tau_{\rm SSE}(M_{\rm WD})$ by linearly interpolating in the mass-formation time relation of isolated stars from SeBa.
We find that the mean difference in WD formation time $\Delta\tau \equiv \tau_{\rm BE} - \tau_{\rm SSE}$ varies  from $-10$Myr to $1.1$Gyr between our models (Figure \ref{fig:dtau_mwd}, see also Table \ref{table:results_table} and Appendix\,\ref{sec:appendix_overview}). However, $\Delta\tau$ can be as large as the assumed age of the Milky Way for individual systems. The absolute difference in WD formation time is larger for the lower-mass end of the WD mass distribution because evolutionary time-scales associated with these stars are larger.
The quantity $\Delta\tau$ can be positive and negative, i.e. binary mergers can form similar mass WDs in both longer and shorter times compared to isolated stars, as we briefly explain in the following.\\ 
Typically, when $\tau_{\rm BE} > \tau_{\rm SSE}$, two lower-mass components form a more massive star with little mass loss preceding and during the merger. Eventually a relatively massive WD is formed, which would have had a shorter formation time if formed by an isolated progenitor \citep[analogous to massive star mergers as core-collapse progenitors][]{DeDon03,Zap17}. Typically, this is the most important effect of BE upon the age difference of single WDs. Typical examples of channels that contribute to this scenario are the WD+WD mergers and MS+MS mergers (see Section \ref{sec:ev}).

Conversely, if the binary components lose a considerable amount of mass before merging, a relatively low-mass single WD is formed. This WD has a long associated SSE time, where as the initial massive binary components 
have a relatively short MS lifetime due to their mass. 
Specifically HG+MS (and secondarily RGB+MS) mergers contribute to the $\tau_{\rm BE} < \tau_{\rm SSE}$ WDs, since in these channels a significant amount of mass can be lost in stable mass transfer preceding the merger (see Section \ref{sec:ev}).

\begin{figure}[h!]
	\centering
    \includegraphics[width=\hsize]{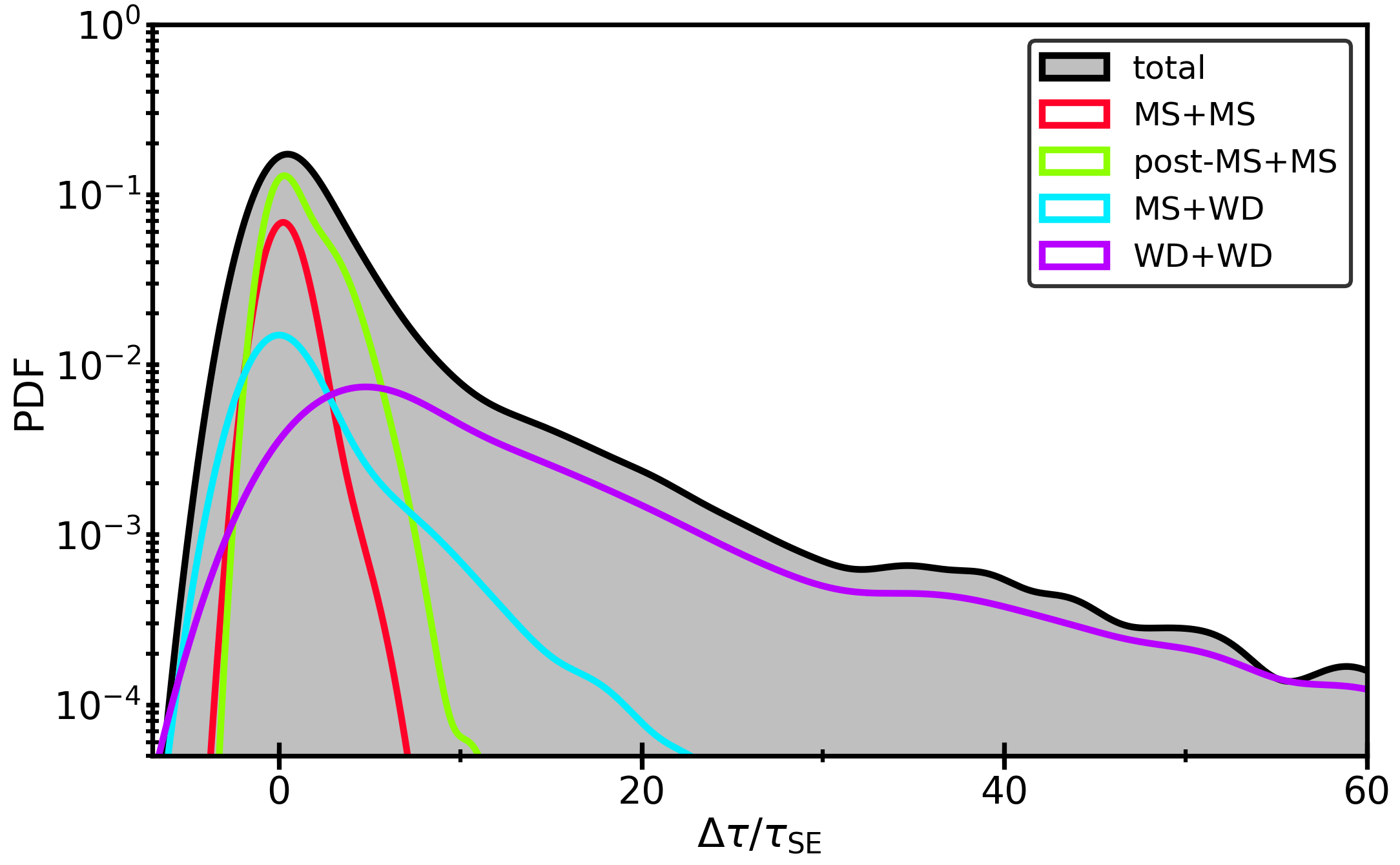}
	\caption{Distributions of WD formation time difference (as fraction of the inferred SSE time) in our default model for single WDs from binary mergers. The shaded histogram is the distribution of all WDs in the sample. The coloured lines indicate various types of mergers, as shown in the legend. The black line and shaded grey area show the PDF for all single WDs formed through binary mergers. This also includes types of merger that are not explicitly shown as a separate coloured line. 
     }
	\label{fig:dtau_age_default}
\end{figure}

In Figure \ref{fig:dtau_age_default}, we show the WD formation time difference relative to the associated SSE time ($\Delta\tau/\tau_{\rm SSE}$). The distributions are shown for our default model, which is qualitatively representative for all models. Whilst shorter evolutionary times are also found, binary interactions mostly work to prolong the single WD formation time. The $\Delta\tau/\tau_{\rm SSE}$ ranges from $\sim -2.5$ up to $\sim60$. For all our models, the mean BE time is longer than the associated SSE time by a factor of around $3-5$ (table \ref{table:results_table}).

We also show the specific contributions from various types of mergers in Figure\,\ref{fig:dtau_age_default}.
Mergers involving two MS stars typically do not result in very large differences in formation times. Larger differences are found  for mergers involving more evolved components. 
The largest difference are found for the DWD mergers, which can have formation times up to several factors of ten times greater than the corresponding SSE time.

\subsubsection{Discrepancies in WD age measurements}
The total age $\tau$ of a WD is given by $\tau = \tau_{\rm form} + \tau_{\rm cool}$, where $\tau_{\rm form}$ is the WD formation time (i.e. $\tau_{\rm BE}$ or $\tau_{\rm SSE}$), and $\tau_{\rm cool}$ is the WD cooling time. Because of the difference in formation times, WDs formed through binary mergers might appear younger or older than their true age, if only SSE is considered. We define the \textit{apparent} age $\tau_{\rm app}$ as $\tau_{\rm app} = \tau_{\rm cool} + \tau_{\rm SSE} (M_{\rm WD})$, where $\tau_{\rm SSE} (M_{\rm WD})$ is inferred from our single stellar model.

\begin{figure}[h!]
	\centering
    \includegraphics[width=\hsize]{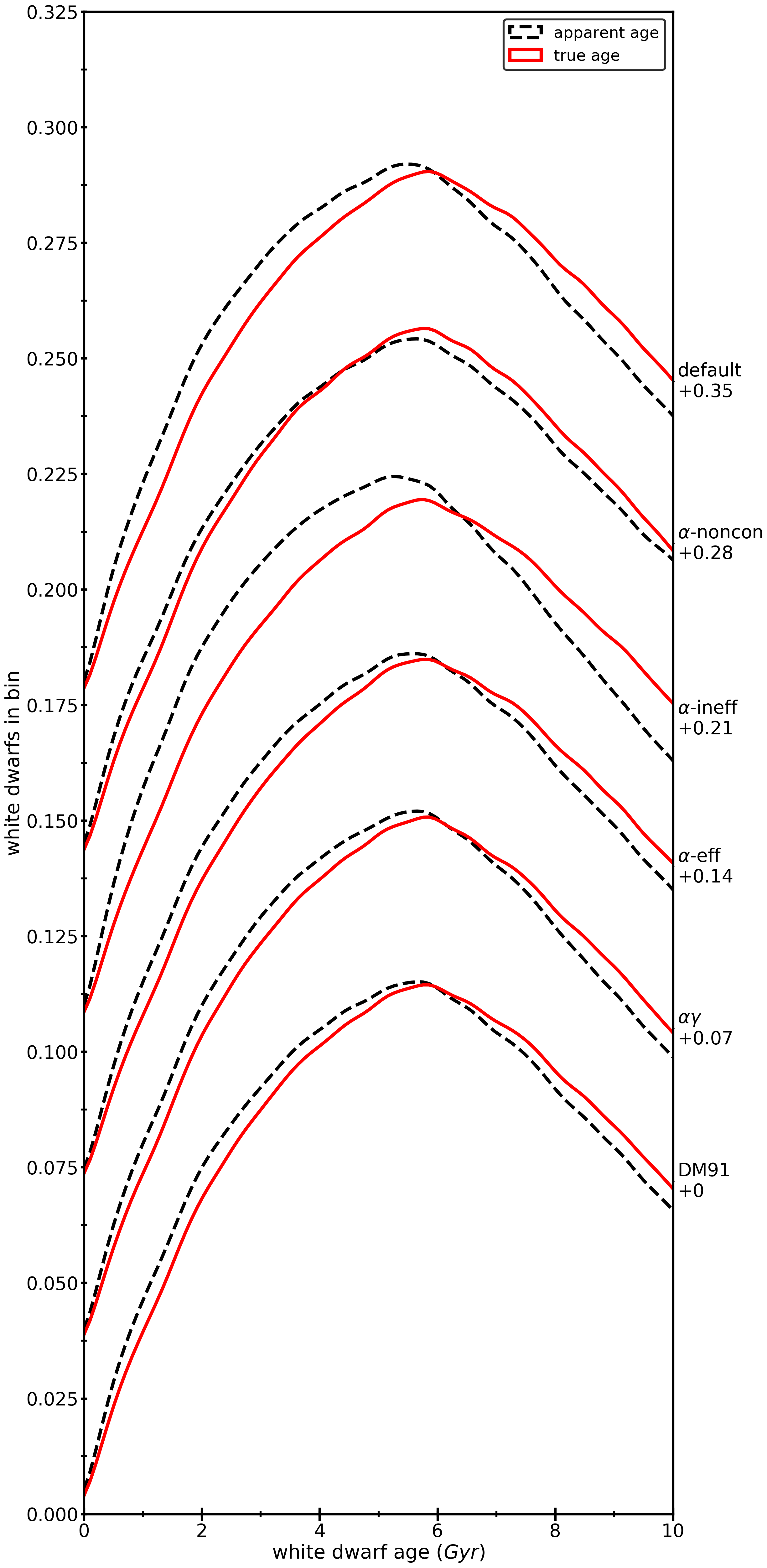}
	\caption{ Distributions of WD ages including single WDs from both binary mergers and isolated single stars. For each model, the dashed black line indicates the apparent age distribution, whilst the solid red line indicates the true age distribution. The distributions shown include an offset (indicated on the right y-axis) for visibility.
	}
	\label{fig:ages_all_models}
\end{figure}

In Figure \ref{fig:ages_all_models} we present the apparent and true age distribution of all single WDs (including single WDs from isolated single stars). Even for the full population, we find an evident difference between the true and apparent WD age distributions: the median ages of WDs appear ${\sim}85-430$Myr smaller than they actually are. The difference between the apparent and true age distribution is mostly affected by the number of WDs that come from mergers. 

The age distributions increase over the first several Gyr, as the common (low-mass) WDs take considerable time to form. The strong decline in the age distributions at later times is  mainly determined by the magnitude limit of the sample which excludes preferentially the old and dim WDs. A detailed study of the dependence of our results on the observational selection effects we introduced is provided in Section\,\ref{sec:effect_of_cuts}. 

\begin{figure}[h!]
	\centering
    \includegraphics[width=\hsize]{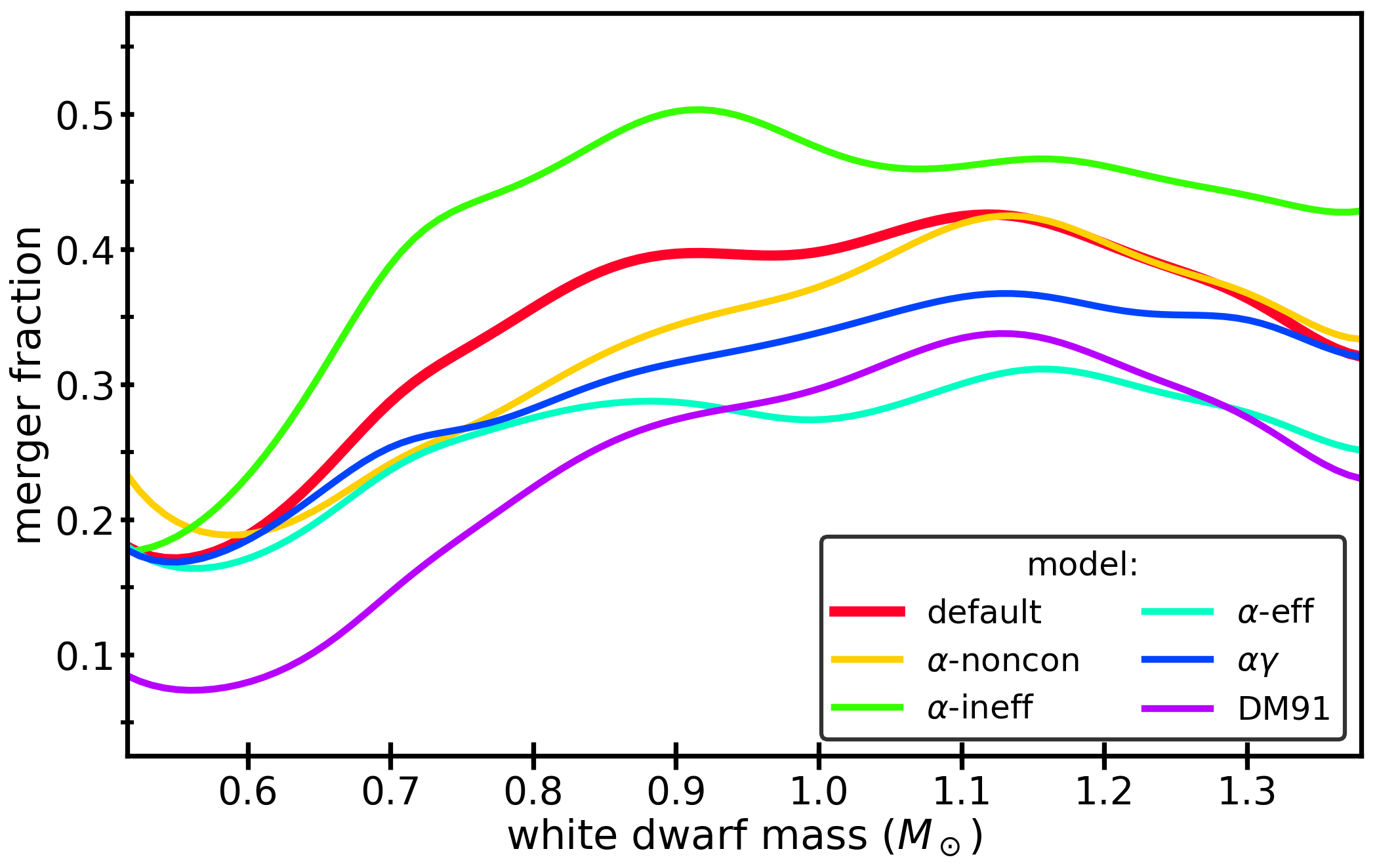}
	\caption{Fraction of WDs in the total population (isolated and binary progenitors) that was formed through binary mergers at each mass. The colours correspond to our different BPS models, as indicated in the legend.}
	\label{fig:mass_fractions}
\end{figure}

\subsection{Impact on WDs as age indicators}
\label{sec:impact}
In the previous section we showed that single WDs with a binary origin affect the predicted age distribution of WDs. 
Overall, the WD formation times would be underestimated by a factor of about 3.1-5 and the median age of the population by about 50-450 Myr. At first glance, stellar mergers give rise to an intrinsic error affecting about 10-30\% of single WDs. This naturally gives rise to the question: What can we do to mitigate this? 

The importance of binary mergers for single WDs generally increases with WD mass. In Figure \ref{fig:mass_fractions}, we present the fraction of binary mergers to the full population of observable single WDs, which we call the `merger fraction'. Additionally, the WD formation time difference between BE and SSE ($\Delta \tau$) varies with WD mass (Figure\,\ref{fig:dtau_mwd} and Table\,\ref{table:results_table}). 
Binaries typically form low-mass WDs faster than expected from SSE, whilst more massive WDs are generally produced in longer times than expected. Combined ,this means that the contamination from mergers to the age distributions differs for different mass ranges.  
Assuming that the SFH is independent of the progenitor mass, this could be used to place constraints on the merger fraction. Inferring WD ages in separate mass bins requires a large sample to avoid small number statistics and a good understanding of WD cooling and other selection effects.

Another way to identify stellar mergers is by comparing the photometric age (as discussed so far) with the dynamical age of the WD. The latter is based on the empirical relation between age and velocity dispersion \citep{Hol09,Weg12, Ang17}. 
Both \cite{Dun15} and \cite{Che19} have applied this method to DQ (massive carbon-rich) WDs to find that their large transverse velocities are inconsistent with a young population resulting from SSE alone. In particular \cite{Che19} find a clear discrepancy between the photometric and dynamical age for $20\pm6\%$ of WDs in the mass range $1.1-1.28\Mo$, and suggest a merger origin. Given that only the strongest outliers can be identified, this is in good agreement with our results in Section\,\ref{sec:massive_wds}.

It has also been suggested that during some merger processes a magnetic field can be generated or amplified \citep{Tou08,Nor11, Gar12,Wic14} that can be used to identify merger remnants. The WDs with high magnetic fields ($B \gtrsim 10^6$G) are generally more massive \citep[e.g.][]{Sil07,Gen18} than average WDs. 
Both \cite{Gar12} and \cite{Bri15} performed a BPS study to evaluate the expected characteristics of magnetic single WDs. The evolutionary channels they consider are a subset of those in our study and are in general agreement with our results for massive WDs (magnetic or non-magnetic), as well as the mass distribution and fraction of mergers.

\begin{figure}[h!]
	\centering
    \includegraphics[width=\hsize]{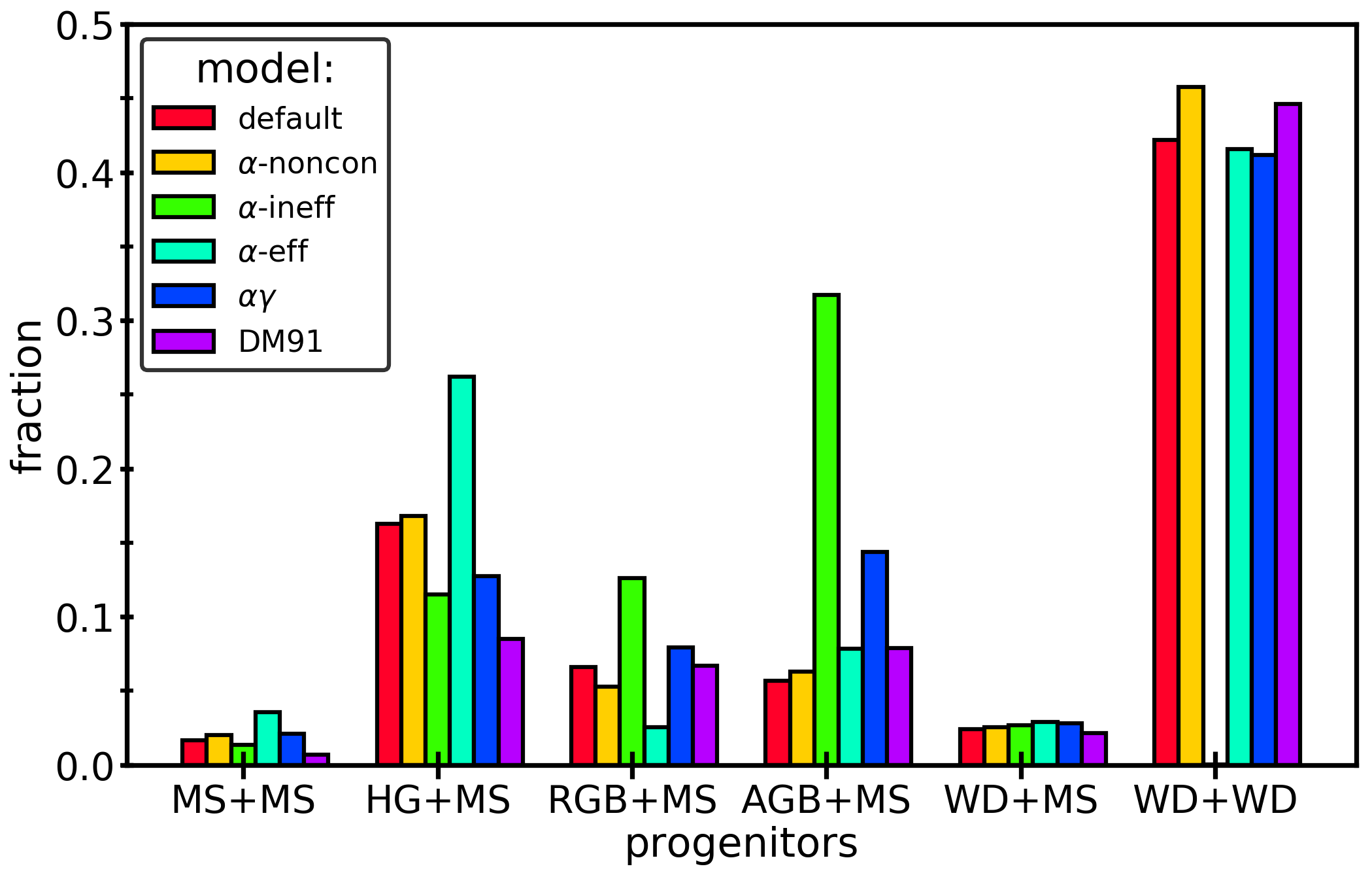}
	\caption{ Fractional contribution of various formation channels, similar to figure \ref{fig:progenitor_contributions_all} but for massive ($M\geq 0.9 M_\odot$) WDs. The height of the bars corresponds to the fraction of massive single WDs that forms through a certain type of merger relative to all WDs from binary mergers. The specific type of merger is indicated on the x-axis.
	}
	\label{fig:massive_merger_types}
\end{figure}

\subsection{Massive white dwarfs}
\label{sec:massive_wds}
Binary mergers are a significant source of massive WDs in all models. For the formation of WDs with masses $\geq 0.9 M_\odot$, which we refer to as `massive' WDs in this work, ${\sim}30-50 \%$ of single WDs form through a binary merger compared to ${\sim}10-30 \%$ for the full mass range.

For massive WDs, different types of mergers are important compared to when the full mass range is considered (Figure \ref{fig:massive_merger_types}). There is almost no contribution from mergers between two MS stars and the RGB+MS mergers are not the dominant type of merger. 
For massive WDs, the most important type of merger is typically WD+WD. In most of our BPS models, these DWD mergers are responsible for about $45\%$ of all the mergers leading to a single WD with mass $\geq 0.9 M_\odot$. The exception is model $\alpha$-ineff, where AGB+MS mergers are the dominant type of merger and responsible for around $35\%$ of all mergers leading to a single massive WD.  

\begin{figure}
    \centering
    \includegraphics[width=\hsize]{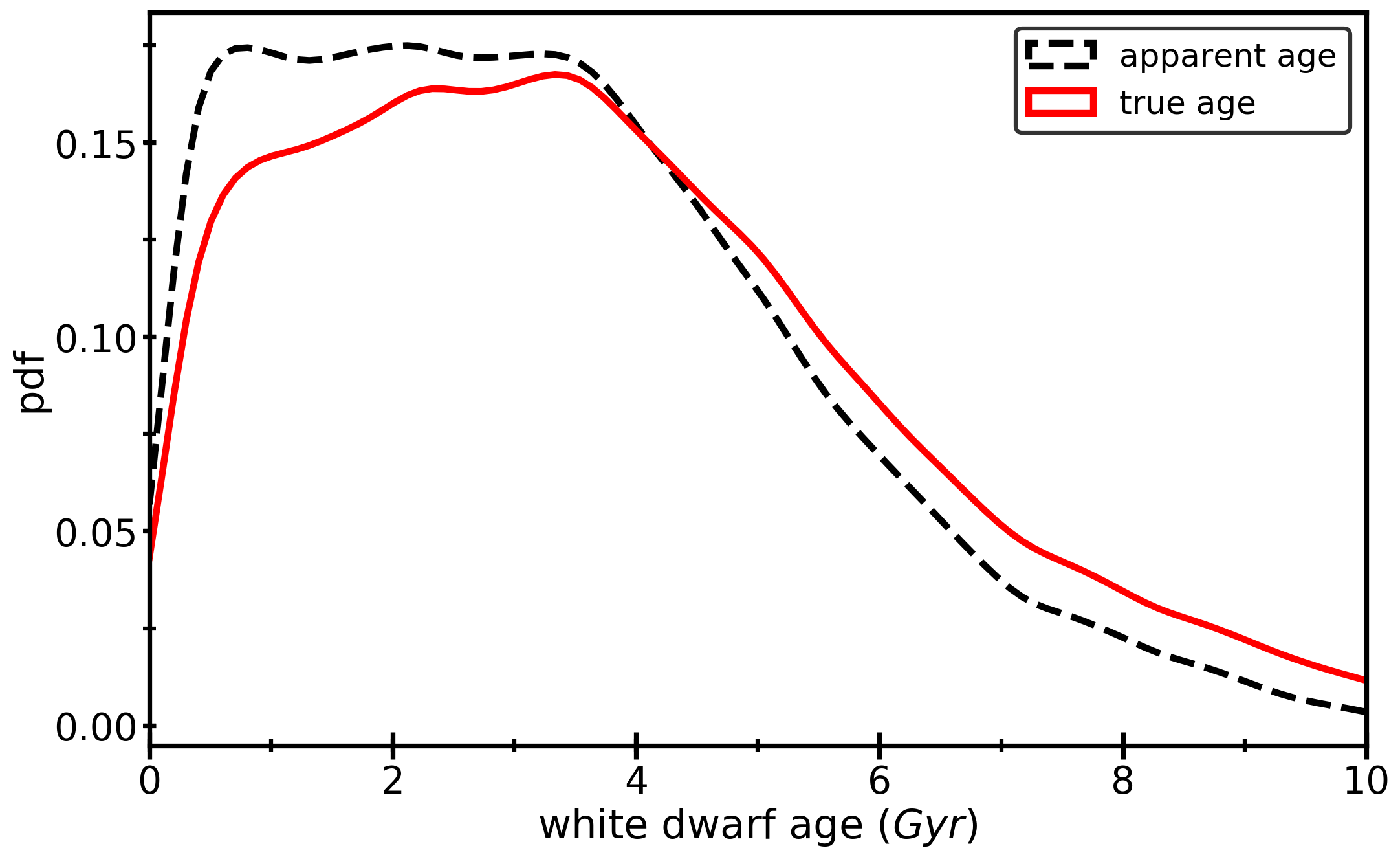}
    \caption{Age distributions of observable single massive ($M > 0.9 M_\odot$) WDs in our default model sample. The true (solid red line) and apparent (dashed black line) age distributions for the total population of single WDs are shown. This population consists of WDs from isolated stars and WDs from our default BPS model.}\label{fig:massive_dtau_age_default}
\end{figure}

The mean formation time of a massive WD from BE is a factor of $5-13$ longer than that from SSE. This is also larger than in the full mass range in which we found factors of $3-5$. Individual massive WDs from binary mergers appear on average $-0.5-1.1$Gyr younger than they truly are.

The age distributions (Figure \ref{fig:massive_dtau_age_default}) show a more rapid decline after the peak compared to that in the full mass range because of the dependence of cooling on WD mass. The median of the distributions of apparent ages is about $200-390$Myr lower than that of the distributions of true ages.

Regarding massive WDs, there has been a long-standing discussion concerning their mass distribution, in particular about an alleged excess of WDs with masses $\sim 0.8M_{\odot}$ \citep[see e.g.][]{Mar97, Ven99}.  The discussion is currently two-fold. Firstly, it is not clear whether or not the excess is statistically significant. Secondly, the origin of the alleged second peak in the mass distribution is debated \citep[inter alia opera][]{Gia12,Kle13,ElB18, Kil18, Reb15,Tre16,Kep16,Hol18, Ber19, Che19}. Amongst other suggestions, several of these studies have suggested a stellar population of merged binary stars as the cause.
Using the second data release from Gaia, \cite{Kil18} studied the mass distribution of single WDs observationally and through a BPS model. 
The authors conclude that a combination of single WDs that evolve in isolation and single WDs that form through mergers in binary systems can naturally explain both the number and mass distribution of the WD sample. In their simulations, the authors find that $\sim 14\%$ of single WDs form through mergers in binary systems, and that the mean WD mass from mergers is $0.74 \pm 0.19 M_\odot$. This is in perfect agreement with our findings for our model DM91, which is the most similar to their model. 
However, all our BPS models, which include several models for BE, primordial populations, WD cooling, and selection effects, lead to a WD mass distribution that is smooth around intermediate masses.
In this region, there is a contribution from binary mergers; the number of WDs varies between the models,  but the general shape of the mass distribution does not. We do not find evidence that binary mergers lead to an excess of $M\sim 0.8M_\odot$ WDs in the observable $100$pc sample.

Alternatively, it has been suggested that the exclusion of cold and dim WDs could result in an excess of WDs at higher masses  \citep{Ise13}. In all our models, we observe no such excess when only WDs with effective temperatures higher than $12,000K$ are considered.

\subsection{Merger rates}
\label{sec:rates}

\begin{figure}[h!]
	\centering
    \includegraphics[width=\hsize]{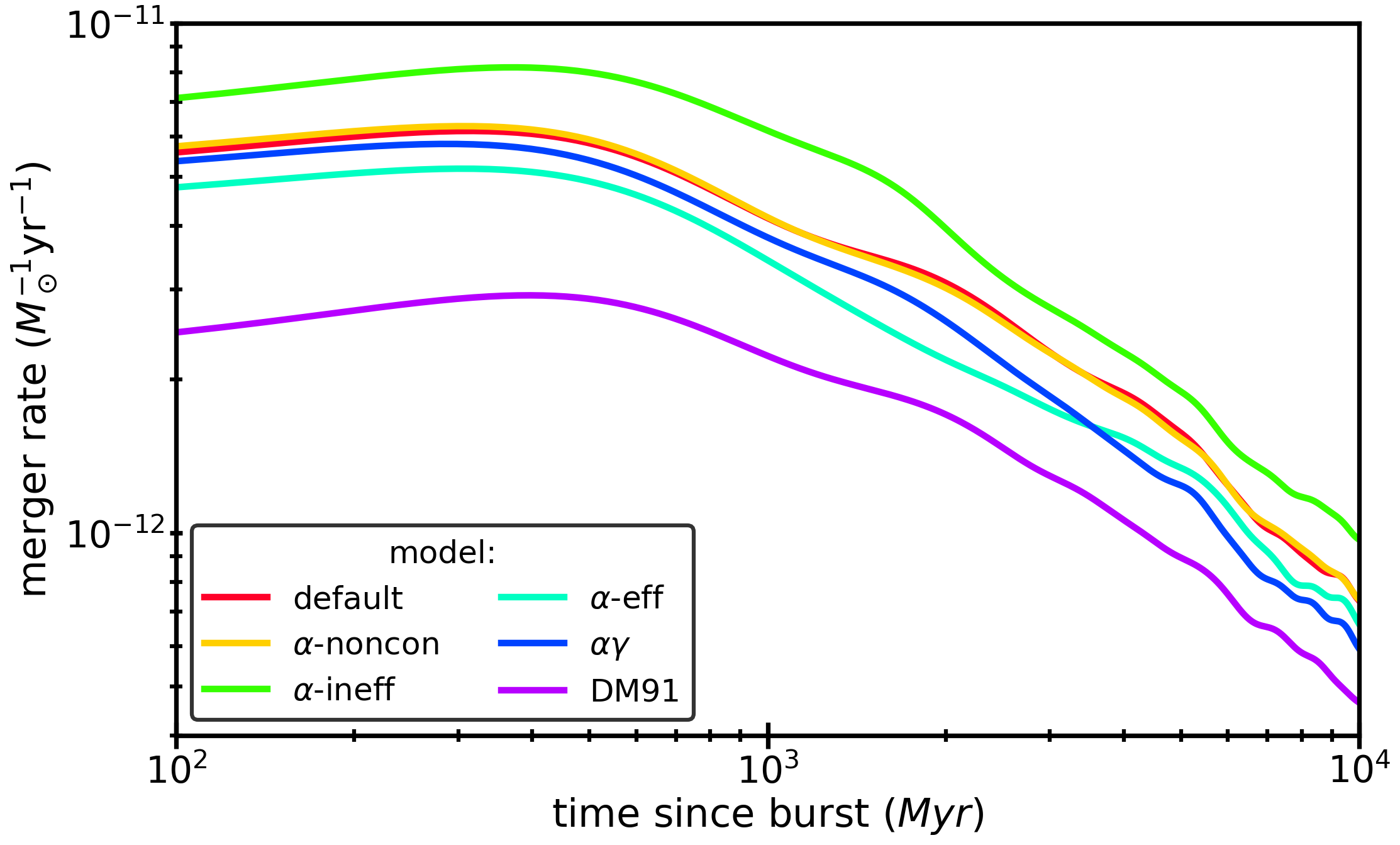}
	\caption{Rates of mergers whose remnant forms a single WD within $10$ Gyr. The rates are given as the predicted number of all merger events per simulated solar mass per year after a burst in star formation at $t = 0$.
	}
	\label{fig:merger_rates}
\end{figure}

Throughout the years, the merger phase in BE has been associated with various classes of transients. Therefore, the observational properties and rates of various types of transients can help to test the binary merger models.  
We present our results regarding the rates of mergers leading to single WDs. Figure \ref{fig:merger_rates} shows the merger rate as a function of time after a burst of star formation at $t=0$. We find that the merger rate has the same shape for most of our models: at relatively early times (up to ${\sim}0.1$ Gyr), the merger rates are dominated by MS+MS mergers. At later times, mergers involving a WD become increasingly dominant, and the total merger rates fall off steeper. In model DM91, there is a lack of mergers at earlier times compared to our other models, because this model does not feature a significant number of MS+MS mergers. The total integrated merger rate is $0.02 - 0.035$ per solar mass of created stars. For the Milky Way this would translate to a rate of $0.06-0.11$ yr$^{-1}$, assuming a constant star formation rate of $3\,M_\odot$ yr$^{-1}$ (see Section \ref{sec:mw}).

Several estimates of the binary merger rate have been made based upon observations. 
For example, based on the blue straggler formation rate of $\sim0.1$yr$^{-1}$ \citep{Cia05}, \cite{Sok06} infer a merger rate of bright transients of ${\sim}0.02 - 0.1$yr$^{-1}$, in general agreement with our estimates.
Luminous red novae (LRNe) form another class of transients that has been linked to mergers \citep[as clearly demonstrated for V1309 Sco,][]{Tyl11}, as well as CE-ejections. The rate of LRNe has been estimated by \cite{Ofe08} and \cite{Koc14}. The former study derived a lower limit of $0.019$yr$^{-1}$ based on two events, and the latter a rate of the order of 0.5yr$^{-1}$ based on a handful of events. 
Comparing the observational estimates of LRNe to our synthetic rates shows that the synthetic rates are somewhat low. This is not surprising because we focus on mergers that lead to single WDs, and therefore do not consider binaries with two low-mass stellar components or binaries that survive a CE phase. Based on a BPS model (similar to our method), \cite{Koc14} estimates a theoretical rate of $0.2$yr$^{-1}$. Overall, within the current uncertainties, the synthetic and observed rates are consistent. 

Lastly, we discuss the predicted merger rates of double WDs.
From the joint-likelihood analysis of WD samples from the Sloan Digital Sky Survey and ESO-VLT Supernova-Ia Progenitor surveY, \cite{Mao18} find that, integrated over the lifetime of the Galaxy, $8.5-11$ per cent of all WDs ever formed have merged with another WD. This is significantly higher than what we found in this work (${\sim} 1-3\%$). The discrepancy can be traced back at least partially to a different fraction of double WDs per single WD. Our models are consistent with the fraction found in the $20$ pc sample of WDs \citep{Too17}, but a factor of 2.5-5 lower than the fraction derived in \cite{Mao18} \citep[see also ][for a discussion]{Reb18}.

 \begin{table*}
\caption{Summary of our main results. 
The first column shows the mean WD formation time difference (that is the mean of $\Delta\tau \equiv \tau_{\rm BE} - \tau_{\rm SSE}(M_{\rm WD})$). 
 Similarly, $\langle \tau_{\rm BE} / \tau_{\rm SSE}\rangle$ indicates the mean ratio of the BE time to the corresponding inferred SSE time. The quantity $\Delta T$ denotes the difference between the median value of the true age and the corresponding apparent age distribution, inferred assuming single stellar progenitor evolution (see also Section \ref{sec:effect_of_cuts}). Lastly, the merger rate is defined as the predicted number of mergers (of any type) leading to a single WD after a burst of star formation $10$Gyr ago per solar mass of created stars. In this table, a quantity with subscript '\textit{m}' indicates that this quantity is calculated for the massive ($M \geq 0.9M_\odot$) WDs in our models.}             
\label{table:results_table}      
\centering      
\begin{tabular}{l|ccccccc}
\toprule
model name      & $\langle \Delta\tau\rangle \; (\text{Gyr})$ & $\langle \Delta\tau\rangle_m \; (\text{Gyr})$ & $\langle \tau_{\rm BE} / \tau_{\rm SSE}\rangle $ & $\langle \tau_{\rm BE} / \tau_{\rm SSE}\rangle _m$ & $\Delta T \; (\text{Gyr})$ & $\Delta T_{m} \; (\text{Gyr})$ & \makecell{Integrated \\ merger rate $(M_\odot^{-1})$} \\ \hline
default         & 0.84                                        & 1.03                                          & 3.88                                            & 11.35                                             & 0.292                                 & 0.388                                   & 0.024                                                 \\
$\alpha$-noncon & -0.01                                       & 0.89                                          & 3.16                                            & 10.34                                             & 0.084                                 & 0.326                                   & 0.024                                                 \\
$\alpha$-ineff  & 1.08                                        & 0.48                                          & 3.13                                            & 4.98                                              & 0.433                                 & 0.207                                   & 0.032                                                 \\
$\alpha$-eff    & 0.56                                        & 1.21                                          & 3.53                                            & 14.0                                              & 0.183                                 & 0.298                                   & 0.020                                                 \\
$\alpha\gamma$  & 0.58                                        & 0.92                                          & 3.13                                            & 10.54                                             & 0.193                                 & 0.293                                   & 0.021                                                 \\
DM91            & 0.98                                        & 1.14                                          & 5.01                                            & 12.49                                             & 0.174                                 & 0.31                                    & 0.013                                                
\end{tabular}
\end{table*}

\section{Discussion}
\label{sec:disc}
Stellar mergers are common. We have demonstrated that this has an impact on the age distribution of single WDs. In the following section, we discuss the uncertainties in and limitations of our modelling approach.

\subsection{Dependence on selection effects}
\label{sec:effect_of_cuts}
In this section we investigate how selection effects influence our synthetic models and the corresponding WD age distributions. 
In particular, we vary the Galactic volume through the limiting distance $d_{\rm lim}$, and the limiting magnitude $G_{\rm lim}$ in the Gaia G band.

With our default assumptions $d_{\rm lim}=100$pc and $G_{\rm lim}=20$, we find ${\sim} 8,600$ WDs formed by SSE, and ${\sim} 1,350 -4,000 $ single WDs formed by binary mergers in the different models. These numbers agree to within a factor $\sim 1.5$ with the 100 pc sample acquired from the second Gaia data release\footnote{The discrepancy can be partially explained by the difference between modelled and observed space density of WDs, which typically agree to within a factor ${\sim}2$ \citep{Too17,Hol18}. }, for which studies have found between $\sim 15000$ \citep{Gen18} and $\sim 18000$ \citep{Jim18}  single WDs in $100$pc. In our samples, we removed $19\%$ of WDs owing to their insufficient brightness, giving a completeness of $81\%$. At the massive end, the completeness decreases to $61\%$. 

Interestingly, WD samples of binary mergers are typically more complete than those from pure SSE. This is because  on average they are  more massive and younger (and therefore brighter) than single WDs formed through SSE. 
This is especially noticeable for the massive WDs, where the completeness levels are $65\%$ versus $59\%$, respectively. If we increase the observational magnitude limit to $G_{\rm lim}=21$ and $22$, only ${\sim}3\%$ and ${\sim}1\%$ of WDs fall below the brightness limit. With the latter limit, the sample remains almost complete (96$\%$) out to $150$pc.

\begin{figure*}[h!]
    \centering
\includegraphics[width=\hsize]{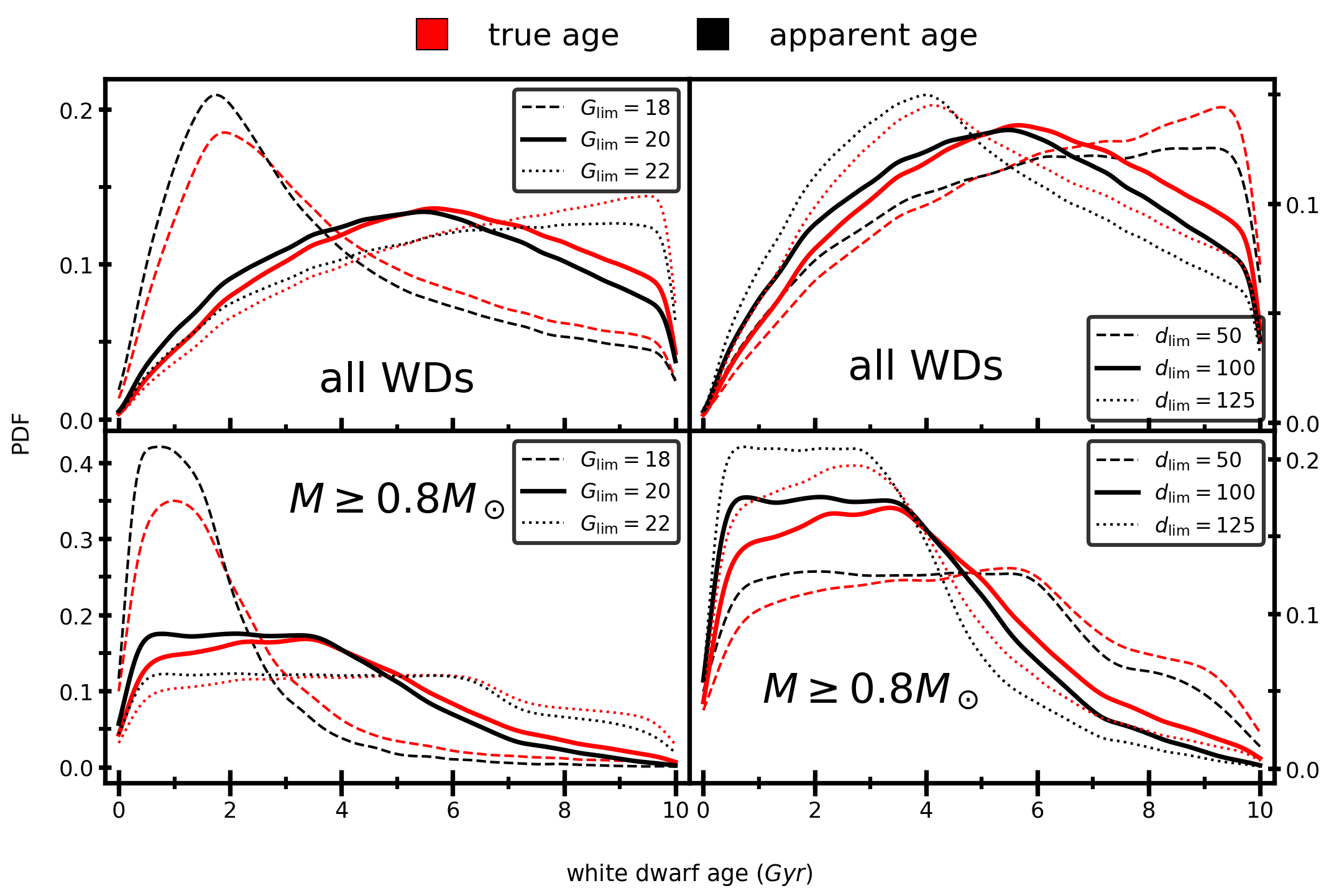}
    \caption{Dependence of our results on observational selection effects and Galactic volume. In all panels, the black lines indicate the apparent age distributions, whilst the red lines indicate the true age distributions, similar to Figures\,\ref{fig:ages_all_models}~and~\ref{fig:massive_dtau_age_default}. The line style indicates either the limiting Gaia G-band magnitude (left column) or the Galactic volume (right column). In the top row, we present results for our full default population of single WDs from both isolated stars and binary mergers; the bottom row contains the same plots, but restricted to the massive ($M \geq 0.9M_\odot$) single WDs.}
    \label{fig:effect_of_cuts2}
\end{figure*}

In Figure \ref{fig:effect_of_cuts2}, we show how the age distributions change when we vary the limiting magnitude. With a deeper survey, more old and cold single WDs would be detected, which affects the tail of the age distribution. For massive WDs, the plateau is extended to older WDs. The difference in the median of the true and apparent age distributions (that is $\Delta T$ in Tbl.\,\ref{table:results_table}) remains roughly constant. At a lower limiting magnitude of $G_{\rm lim}=18$, the age distribution is strongly affected by the selection effects (more so for the massive WDs), but the age discrepancy remains present. 

In Figure \ref{fig:effect_of_cuts2}, we also show the age distributions for WD samples of different Galactic volumes at constant $G_{\rm lim}=20$. A 50\,pc sample, with its high level of completeness, resolves better the population of old and cold WDs. The offset between the apparent and true ages is similar to that of the $100$ pc sample. Extending the sample to larger distances reduces the completeness of the samples, and increases the difference in the median of the true and apparent age distributions. The completeness is typically larger for the sample of massive single WDs compared to the full mass range out to volumes of $d\lesssim 150$pc.

Most importantly, Figure \ref{fig:effect_of_cuts2} affirms that, although sensitive to the sample definition, there is always a noticeable difference between the true and apparent ages of single WDs, especially for the massive single WDs.

\subsection{Limitations of the model}
\label{sec:limitations}
In this section, we critically examine additional sources of uncertainty in our results. For the current study, we used the population synthesis approach to model the impact of merger remnants on the population of single WDs. This method is appropriate for investigating the demographics of stellar populations given our current understanding of SSE and BE. In order to reduce the computational expense, a number of simplifying assumptions are typically  made. In most binary population synthesis studies, stellar evolution is simulated in a parametrised way, as is the case in this work. 
We approximate the evolution of the merger remnant by finding the best fitting SSE track (Section\,\ref{sec:mergers}), which is calculated for single stars in hydrostatic and thermal equilibrium. A detailed modelling of the merger process and remnant is beyond the scope of this project and merits further investigation.

One of the uncertainties in our modelling is how much mass is lost during the merger process and the phases directly preceding and following it. Our default model is that the mergers take place conservatively (as roughly expected for MS+MS or WD+WD). To test the effect of this assumption, one of the additional models ($\alpha$-noncon) considers partial envelope ejection during a CE phase in proportion to the available orbital and required energy to the unbind the envelope (see Eq.\,\ref{eq:noncon}). As shown in Fig.\,\ref{fig:ages_all_models} (top two lines), the apparent and true age distributions are dependent on the level of conservativeness we apply in the simulations, but both models show a clear distinction between the apparent and true age distribution. 

Another assumption that could affect the merger remnants is the modelling of stable mass transfer. In this context, we specifically mention mergers between a HG star and a MS companion. 
This channel contributes about $7.5-12.5\%$ of the mergers leading to a WD (see Fig.\,\ref{fig:progenitor_contributions_all}) and they occupy the $\Delta \tau \leq 0$ region in Figure \ref{fig:dtau_mwd}. In these systems, the mean amount of mass lost before the merger is $20-30\%$ of the initial total mass in the systems. This mass is generally lost during the phase of stable mass transfer directly preceding the fatal CE phase. 

For the merger of a double WD, we have made two simplifying assumptions. First, the merger of He WD and a CO WD (or two massive He WDs) can lead to re-ignition of helium and the formation of an extreme helium
star \citep[e.g.][]{Jus11, Hal16} or an R Coronae Borealis star \citep{Web84,Ibe96}. In this work we do not take into account this additional phase of evolution and the additional time this nuclear burning phase adds to the precursor lifetime of the WD. It is expected to be of the order of 0.1Myr \citep{Zha12, Zha14,Men13}, where the mean age difference $\Delta\tau$ for the He WD + CO WD mergers is ${\sim} 2-3 $Gyr in our models. 
Potential wind mass loss \citep[e.g.][]{Sch19} can affect the apparent age, and reduce the difference between the apparent and true age for those WDs resulting from a R Coronae Borealis star.   

Secondly, we assume that the cooling age of the remnant WD is not affected. 
In other words,  we assume that after the merger a hot WD is formed that cools in a similar fashion as a normal WD. However, merger remnants might cool at a different rate, depending on the details of the merger process, and whether their internal and atmospheric compositions are different. Moreover, detailed simulations of WD mergers show that typically the less massive WD is disrupted in the merger. Its material builds up around the remaining WD \citep{Gue04, Yoo07, Lor09, Pak11}, whose core temperature is not affected significantly, assuming core burning is avoided. An exception to this occurs when the initial WDs have comparable masses and the merger leads to a complete disruption of both components. In that case the central region of the merger remnant is heated by viscous heating. 
In order to assess the effect that this cold core could have on our results, we re-calculated the cooling of the WDs resulting from DWD mergers. Instead of assuming the merger remnants start cooling from the same temperature as a WD formed by SE, we assume that they begin cooling at the core temperature of the most massive WD just prior to the merger.  Assuming the cold merger scenario for our default model, we find that $\sim 33\%$ of the WDs formed by DWD mergers would no longer be observable in our $100$pc sample. 
For the full population of single WDs, we find that the cold merger assumption decreases the difference in mean age $\Delta T$ only marginally, by about  $13\%$. For massive WDs, the effect is more significant, since DWD mergers are more abundant in this regime. The difference in mean age $\Delta T$ decreases by about $34\%$ for remnant masses $\geq 0.9 M_\odot$.
We note that in nearby samples, the effect of cooling is less important. For example, in the $50$ pc sample, we only lose around $15\%$ of all DWD mergers. 
 This results in a decrease in $\Delta T$ by $\sim 7.5 \%$  ( and $15\%$ for remnant masses $\geq 0.9 M_\odot$). 
We thus conclude that the details of the merger process, as far as remnant temperature and subsequent cooling is concerned, do not significantly influence our main conclusions. 

Our results are based on a specific IFMR, however different models exist \citep[see e.g.][and references therein]{Cum18, ElB18}. At low-to-intermediate masses $\lesssim 4 M_\odot$, our IFMR is consistent with these IFMRs within their uncertainties. However, at higher masses there is diversity amongst the different IMFRs. This impacts the number of massive WDs directly and possibly the fraction of mergers at a given WD mass as well. However, it might also be the case that the observationally based IMFR measurements are affected by binary mergers. 

In this study we assumed a constant binary fraction of $50\%$. This is appropriate for A- to G-type stars \citep{Duc13,Rag10, derosa2014}, however the binary fraction generally increases for larger stellar masses \citep[see e.g.][]{Rag10,Duc13,Moe17}. For stars at the upper end of the mass range ($M\sim 7M_\odot$), the binary fraction could be as large as ${\sim} 70\%$.
Typically, the merger remnants from binaries with initially massive primaries form massive WDs (see e.g. Fig.\,\ref{fig:IFMR_comp}), and therefore the contribution of mergers to massive single WDs can be even higher than found in our study.

\section{Conclusions}

\label{sec:concl}
In this work, we performed an extensive population study focussed on currently observable single WDs. We modelled the evolution of isolated and binary stars using the population synthesis code \texttt{SeBa}. For the binaries, we focussed on systems that merge and lead to an observable single WD at some point afterwards. We employed six different BPS models, which include different treatments of the CE phase and the primordial binary populations. We also take into account a star formation history, observational selection effects, and WD cooling to construct six synthetic WD populations for the Milky Way. 

Our main conclusions can be summarised as follows:
\begin{itemize}
	\item We find that the number of single WDs that are formed by merger events in binary systems is significant. Between about $ 10-30\%$ of all observable single WD are formed through binary mergers, depending on our model (Tbl.\,\ref{table:summary_nrs}). This is consistent with the relatively low observed binary fraction of WDs compared to their progenitor stars, and with the observed transient rates of merger-related events. The main sources of uncertainty in this range are the properties of the primordial binaries.
	\item Typically, WDs from binary mergers have a formation time that is longer than the formation time of an equal mass WD formed through SSE. On average, the WD formation through the binary merger channel takes about three to five times longer than if the star would have been a single star for its entire life (Fig.\,\ref{fig:dtau_mwd}~and\ref{fig:dtau_age_default}).
	\item As a consequence, we find a significant difference in the age distribution of single WDs between a population of inherently single stars and a population that includes binary mergers. In other words, if a WD sample were used to measure the age distribution or star formation history of a stellar population, the median age of the WDs would be underestimated. In our simulations which assume a simple constant star formation history, the median age would be underestimated by approximately $80-430$Myr. 
	\item 
	In Section\,\ref{sec:impact} we discuss strategies for mitigating the effect of mergers on the WD age distributions. 
	These include deriving the age distributions for different WD mass bins, as the affect and importance of mergers differs with WD mass, as well as including dynamical age measurements.

\end{itemize}

Furthermore, we find that
\begin{itemize}	
    \item Over the years, there has been an active discussion concerning an alleged excess of WDs with masses ${\sim}0.8M_\odot$ (see \ref{sec:massive_wds}). The presence and origin of this excess is still debated. 
	We address this issue and, more specifically, the suggestion that this excess might be caused by a population of binary mergers. We find that the predicted mass distributions, which include WDs from both isolated stars and binary mergers, decline smoothly with increasing WD mass. Although there is a contribution from binary mergers, we find no evidence for a distinct bump around ${\sim}0.8M_\odot$ WDs due to binary mergers. If the existence of this bump is confirmed \citep[but see][]{Ber19}, the origin likely lies elsewhere, such as in the IFMR \citep[as suggested in][for example]{ElB18}. 
	\item Typically the binary mergers that lead to a single WD involve the merger between a post-MS star with a MS companion. These comprise approximately $30-65\%$ of all mergers leading to a single WD. Mergers between two MS stars are also relevant, and account for about $10-25 \%$ of all binary mergers. Double WD mergers are not the main channel in our models. They are responsible for up to $15 \%$ of all binary mergers leading to a single WD.
	\item For massive WDs, merger events play an even more important role (Fig.\,\ref{fig:mass_fractions}). For WD masses of $M>0.9M_\odot$, $30-45\%$ of all observable single WDs within $100$pc are formed through binary mergers. In this work, the dominant type of merger is a WD+WD merger. 
	For these mergers, the difference in the formation time between the binary merger channel and the corresponding SSE channel is larger than for the full mass range. The true WD age can be up to a factor of a few tens larger than the apparent age assuming SSE. 
	    \item The limiting magnitude and distance assumed for the WD sample do not affect the derived mass distributions. The difference between the true and apparent age distributions is weakly dependent on the magnitude limit and distance cut. Generally, the magnitude of the difference increases when  the limiting distance increased at a given limiting magnitude. However, the effect of binary mergers is observable for a broad range of observational selection methods.	
\item The integrated rate of mergers leading to an observable single WD is $0.013 - 0.031 M_\odot^{-1}$ in our models, which is equivalent to a Galactic rate of $0.04-0.094$yr$^{-1}$. 
\end{itemize} 

In short: Assuming SSE for inferring properties of single WDs gives rise to intrinsic errors as single WDs can also be formed following a binary merger. This is a robust effect that we find in all of our models.

\begin{acknowledgements}
The authors are grateful to Onno Pols for helpful comments and general discussions. We thank Walter van Rossem for helpful information on asteroseismology. We also thank Athira Menon and Josiah Schwab for informative discussions about HeWD-COWD mergers and RCB stars, Nadia Blagorodnova for discussions regarding merger-related transients, and Jos\'e Versteeg-Veltkamp for useful feedback. KDT acknowledges support from NOVA. ST acknowledges support from the Netherlands Research Council NWO (grant VENI [nr. 639.041.645]). EZ  acknowledges support from the Federal Commission for Scholarships for Foreign Students for the Swiss Government Excellence Scholarship (ESKAS No. 2019.0091) for the academic year 2019-2020. BTG was supported by the UK STFC grant ST/P000495. This work benefitted from a workshop held at DARK in July 2019 that was funded by the Danish National Research Foundation (DNRF132).
\end{acknowledgements}

\bibliographystyle{aa}
\bibliography{bibtex_wd}

\begin{appendix} 
\section{Difference in WD formation times}
\label{sec:appendix_overview}

\begin{figure*}[h!]
	\centering
    \includegraphics[width=\hsize]{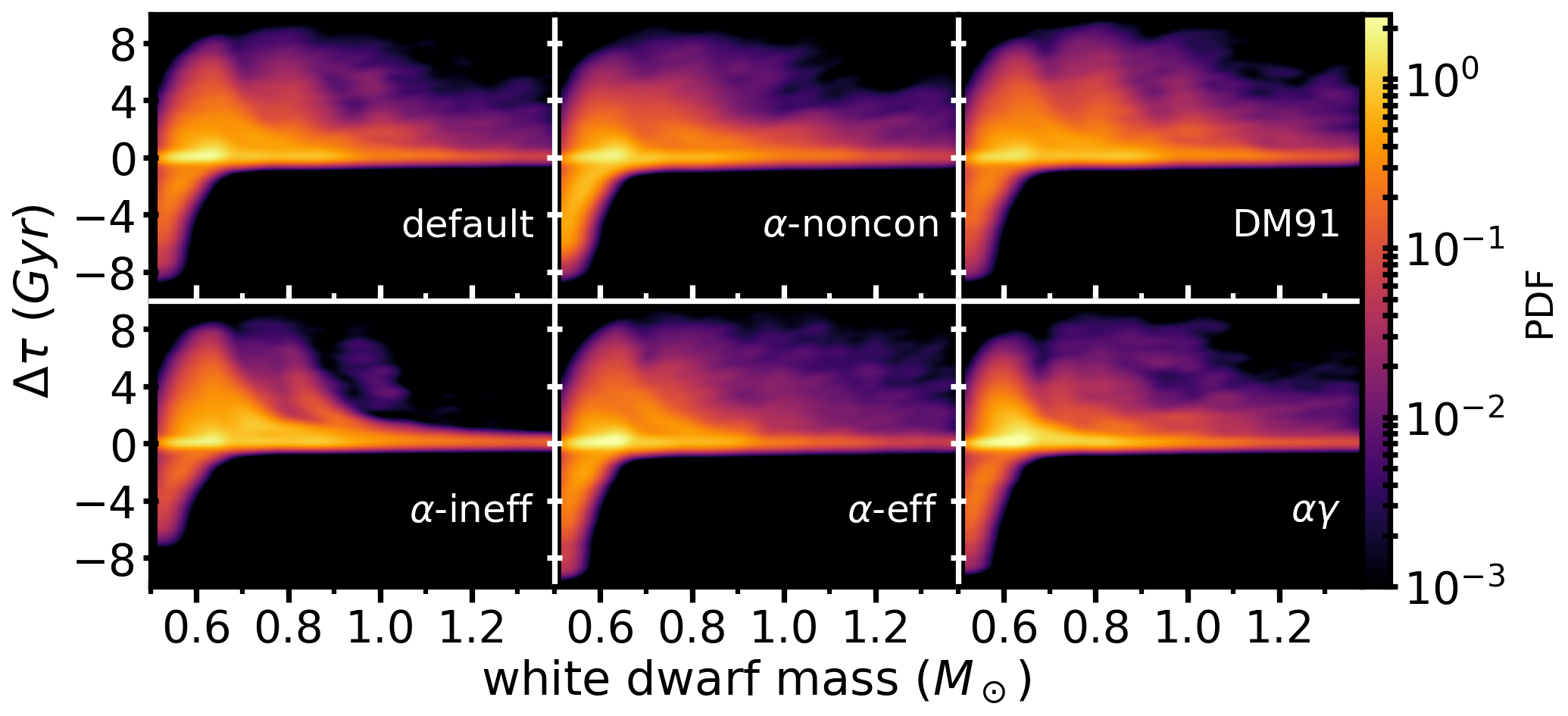}
	\caption{Differences in WD formation time $\Delta\tau$ as a function of WD mass, similar to the top panel of fig \ref{fig:dtau_mwd}. Each panel shows our results for a different BPS models. The colour logarithmically indicates the number of WDs in a given bin, normalised to the total number of WDs from binary mergers. Each panel corresponds to a different BPS model, as indicated at the bottom right of the panels. 
	}
	\label{fig:dtau_mwd_all}
\end{figure*}

In Figure\,\ref{fig:dtau_mwd_all} we show the differences in WD formation time $\Delta\tau$ for all BPS models. The different models result in similar distributions. The largest differences are found in the number of single WDs produced (Section \ref{sec:results}, Tbl.\,\ref{table:summary_nrs}). Smaller differences is the  $\Delta\tau$ distribution arise, and their causes are discussed below. 

The difference in WD formation time depends upon how conservatively the CE phase is modelled (see the top row in Figure \ref{fig:dtau_mwd_all}). In model $\alpha$-noncon , the binaries can lose mass during a CE phase. This additional loss of mass leads to more low-mass WDs, for which $\tau_{\rm SSW}(M_{\rm WD})$ is large. Thus, the binary merger formation time of such WDs is increased with respect to the associated SSE time, resulting in more WDs for which $\Delta\tau < 0$. 

Assumptions regarding the efficiency of the CE phase affect the formation time difference (see the middle row in Figure \ref{fig:dtau_mwd_all}). For example, assuming an inefficient CE phase, ${\sim}80\%$ of the binary mergers occurs through a CE, whereas this fraction is ${\sim}60\%$ in our default model. Since no mass is lost during a CE in these models, 
there is a relative increase in systems with $\Delta\tau > 0$ compared to our default model. 
Sampling the initial orbits according to a log-normal distribution, as in model DM91, results in more WDs located in the two enhancements in $\Delta\tau$ at $M \sim 0.8, 1M_\odot$. These WDs typically form through mergers involving at least one WD, and are more abundant in this model compared to the default model. 

\end{appendix}

\end{document}